/pdfoutput=1
\PassOptionsToPackage{hyphens}{url}
\documentclass[sigplan,10pt]{acmart}
\renewcommand\footnotetextcopyrightpermission[1]{}

\usepackage{caption}
\usepackage{times}
\usepackage{oneinchmargins}
\usepackage{tightenum}

\usepackage{subfig}
\usepackage[normalem]{ulem}

\usepackage{setspace}
\usepackage{epsfig}
\usepackage{graphicx}
\usepackage{times}
\usepackage{amsmath}
\usepackage[compact]{titlesec}
\usepackage{url}
\usepackage{multirow}

\usepackage{tikz}
\usetikzlibrary{calc}
\newcommand*\circled[1]{\tikz[baseline=-3pt]{
            \node[shape=circle,draw,inner sep=1pt,minimum size=10pt] (char) {\small #1};}}

  \renewenvironment{thebibliography}[1]{%
    \begin{oldthebibliography}{#1}%
      \setlength{\parskip}{0ex}%
      \setlength{\itemsep}{0ex}%
  }%
  {%
    \end{oldthebibliography}%
  }

\usepackage{listings}
  \usepackage{courier}
\renewcommand{\em}{\it}

\newcommand{\x}{$\times$}

\newcommand{\ignore}[1]{}

\newcommand{\ulinepara}[1]{\noindent{\uline{\textit{\textbf{#1  }}}}}

\def\cfigure[#1,#2,#3]{
\begin{figure}
\vspace*{0mm}
\begin{center}

\includegraphics[width=3in]{#1} 
 
\vspace*{-3mm}\caption[]{#2
} \label{#3}
 
\vspace*{-5mm}
\end{center}
\end{figure}}

\def\cfigurefour[#1,#2,#3]{
\begin{figure}
\vspace*{0mm}
\begin{center}

\includegraphics[width=4in]{#1} 
 
\vspace*{-3mm}\caption[]{#2
} \label{#3}
 
\vspace*{-5mm}
\end{center}
\end{figure}}

\def\cfiguretemp[#1,#2,#3]{
\begin{figure}
\vspace*{0mm}
\begin{center}

\includegraphics[width=3.5in]{#1} 
 
\vspace*{-3mm}\caption[]{#2
} \label{#3}
 
\vspace*{-5mm}
\end{center}
\vspace*{-2mm}
\end{figure}}

\def\wfigure[#1,#2,#3]{
\begin{figure*}
\vspace*{0mm}
\begin{center}
 \includegraphics[width=\textwidth]{#1} 
 \vspace*{-3mm}\caption[]{#2
} \label{#3}
 
\end{center}
\end{figure*}}

\def\threefigure[#1,#2,#3,#4,#5]{
\begin{figure*}
\vspace*{0mm}
\begin{center}

\begin{tabular}{ccc}
\includegraphics[width=2in]{#1} & \includegraphics[width=2in]{#2} &  \includegraphics[width=2in]{#3} \\
(a) & (b) & (c) \\
\end{tabular}

\vspace*{-3mm}\caption[]{#4
} \label{#5}

\vspace*{-5mm}
\end{center}
\vspace*{-2mm}
\end{figure*}}

\def\dcfigure[#1,#2,#3,#4,#5,#6]{
{
\begin{figure*}
\begin{center}
\begin{minipage}[c]{\columnwidth}{
\includegraphics[width=\columnwidth]{#1} 
\vspace*{0mm}\caption[]{#2} \label{#3} \
}\end{minipage}\hspace*{\columnsep}\
\begin{minipage}[c]{\columnwidth}{
\includegraphics[width=\columnwidth]{#4} 
\vspace*{0mm}\caption[]{#5}\label{#6} \
}\end{minipage}
\end{center}
\end{figure*}
}
}

\def\tableByTable[#1,#2,#3,#4,#5,#6]{
{
\begin{table*}
\begin{center}
\begin{minipage}[c]{3in}{
\centering
{#1}
\vspace*{0mm}\tabcaption[]{#2}\label{#3} \
}\end{minipage}\hspace*{\columnsep}\
\begin{minipage}[c]{3in}{
\centering
{#4}
\vspace*{0mm}\tabcaption[]{#5}\label{#6} \
}\end{minipage}
\end{center}
\end{table*}
}
}

\def\figureByTable[#1,#2,#3,#4,#5,#6]{
{
\begin{figure*}
\begin{center}
\begin{minipage}[c]{3in}{
\centering
\includegraphics[width=\textwidth]{#1}
\vspace*{0mm}\figcaption[]{#2} \label{#3} \
}\end{minipage}\hspace*{\columnsep}\
\begin{minipage}[c]{3.3in}{
\centering
{#4}
\vspace*{0mm}\tabcaption[]{#5}\label{#6} \
}\end{minipage}
\end{center}
\end{figure*}
}
}

\def\tableByFigure[#1,#2,#3,#4,#5,#6]{
{
\begin{figure*}
\begin{center}
\begin{minipage}[c]{4.3in}{
\centering
{#1}
\vspace*{0mm}\tabcaption[]{#2} \label{#3} \
}\end{minipage}\hspace*{\columnsep}\
\begin{minipage}[c]{2.2in}{
\centering
\includegraphics[width=\textwidth]{#4}
\vspace*{-0.35in}\caption[]{#5}\label{#6} \
}\end{minipage}
\end{center}
\end{figure*}
}
}

\def\doublecfigure[#1,#2,#3,#4]{
{
\begin{figure}
\begin{center}
\begin{minipage}[c]{1.5in}{
\begin{center}
\includegraphics[width=1.5in]{#1}
\end{center}
}\end{minipage}\hspace*{1em}\
\begin{minipage}[c]{1.5in}{
\begin{center}
\includegraphics[width=1.5in]{#2}
\end{center}
}\end{minipage}
\vspace*{0mm}\caption[]{#3} \label{#4} \
\end{center}
\end{figure}
}
}

\def\qcfigure[#1,#2,#3,#4,#5,#6]{
{
\begin{figure*}
\vspace*{0.2in}\
\begin{center}
\begin{minipage}[c]{3in}{
\includegraphics[width=3in]{#1} 
\vspace*{-3mm}
}
\end{minipage}\hspace*{0.5in}\
\begin{minipage}[c]{3in}{
\includegraphics[width=3in]{#2} 
\vspace*{-3mm}
}\end{minipage}

\begin{minipage}[c]{3in}{
\includegraphics[width=3in]{#3} 
\vspace*{-3mm}
}
\end{minipage}\hspace*{0.5in}\
\begin{minipage}[c]{3in}{
\includegraphics[width=3in]{#4} 
\vspace*{-3mm}
}\end{minipage}
\end{center}
\caption[]{#5}\label{#6}
\end{figure*}
}
}

\def\twfigure[#1,#2,#3,#4,#5]{
{
\begin{figure*}
\vspace*{0.2in}\
\begin{center}
\begin{minipage}[c]{6.5in}{
\includegraphics[width=6.5in]{#1} 
\vspace*{-3mm}
}
\end{minipage}

\begin{minipage}[c]{6.5in}{
\includegraphics[width=6.5in]{#2} 
\vspace*{-3mm}
}\end{minipage}

\begin{minipage}[c]{6.5in}{
\includegraphics[width=6.5in]{#3} 
\vspace*{-3mm}
}
\end{minipage}
\end{center}
\caption[]{#4}\label{#5}
\end{figure*}
}
}

\def\dwfigure[#1,#2,#3,#4]{
{
\begin{figure*}
\vspace*{0.2in}\
\begin{center}
\begin{minipage}[c]{6.5in}{
\includegraphics[width=6.5in]{#1} 
\vspace*{-3mm}
}
\end{minipage}

\begin{minipage}[c]{6.5in}{
\includegraphics[width=6.5in]{#2} 
\vspace*{-3mm}
}\end{minipage}

\end{center}
\caption[]{#3}\label{#4}
\end{figure*}
}
}

\def\dssfigure[#1,#2,#3,#4,#5,#6]{
{
\begin{figure*}
\vspace*{0.2in}\
\begin{center}
\begin{minipage}[c]{4in}{
\includegraphics[width=4in]{#1}
\vspace*{-3mm}\caption[]{#2} \label{#3} \
}\end{minipage}\hspace*{0.5in}\
\begin{minipage}[c]{2in}{
\includegraphics[width=2in]{#4}
\vspace*{-3mm}\caption[]{#5}\label{#6} \
}\end{minipage}
\end{center}
\vspace*{-0.4in}\
\end{figure*}
}
}

\def\dsfigure[#1,#2,#3,#4,#5,#6]{
{
\begin{figure*}
\vspace*{0.2in}\
\begin{center}
\begin{minipage}[c]{3in}{
\includegraphics[width=3in]{#1}
\vspace*{-3mm}\caption[]{#2} \label{#3} \
}\end{minipage}\hspace*{0.5in}\
\begin{minipage}[c]{3in}{
\hspace*{0.5in}\
\includegraphics[height=3in]{#4}
\vspace*{-3mm}\caption[]{#5}\label{#6} \
}\end{minipage}
\end{center}
\vspace*{-0.4in}\
\end{figure*}
}
}

\def\dsyfigure[#1,#2,#3,#4,#5,#6]{
{
\begin{figure*}
\vspace*{0.2in}\
\begin{center}
\begin{minipage}[c]{2.5in}{
\includegraphics[height=2.5in]{#1}
\vspace*{-3mm}\caption[]{#2} \label{#3} \
}\end{minipage}\hspace*{0.5in}\
\begin{minipage}[c]{2.5in}{
\includegraphics[height=2.5in]{#4}
\vspace*{-3mm}\caption[]{#5}\label{#6} \
}\end{minipage}
\end{center}
\vspace*{-0.4in}\
\end{figure*}
}
}

\def\dyfigure[#1,#2,#3,#4,#5,#6]{
{
\begin{figure*}
\vspace*{0.2in}\
\begin{center}
\begin{minipage}[c]{3in}{
\includegraphics[height=3in]{#1} 
\vspace*{-3mm}\caption[]{#2} \label{#3} \
}\end{minipage}\hspace*{0.5in}\
\begin{minipage}[c]{3in}{
\includegraphics[height=3in]{#4} 
\vspace*{-3mm}\caption[]{#5}\label{#6} \
}\end{minipage}
\end{center}
\vspace*{-0.4in}\
\end{figure*}
}
}

\def\dyoldfigure[#1,#2,#3,#4,#5,#6]{
{
\begin{figure*}
\vspace*{0.2in}\
\begin{center}
\begin{minipage}[c]{3in}{
\epsfysize=2.0in\
\hspace{0.5in}\
\epsfbox{#1}
\vspace*{-3mm}\caption[]{#2} \label{#3} \
}\end{minipage}\hspace*{0.25in}\
\begin{minipage}[c]{3in}{
\epsfysize=2.0in\
\hspace{0.5in}\
\epsfbox{#4}
\vspace*{-3mm}\caption[]{#5}\label{#6} \
}\end{minipage}
\end{center}
\vspace*{-0.4in}\
\end{figure*}
}
}

\def\cfiguredouble[#1,#2,#3,#4]{
\begin{figure}
\vspace*{0.2in}\
\begin{center}
\begin{minipage}[c]{1.5in}{
\epsfxsize=1.5in\
\epsfbox{#1}
}\end{minipage}\hspace*{0.1in}\
\begin{minipage}[c]{1.5in}{
\epsfxsize=1.5in\
\vspace{0.1in}\epsfbox{#2}
}\end{minipage}\vspace*{-0.10in} \caption[]{#3}\label{#4}
\end{center}
\vspace*{-0.4in}\
\end{figure}
}

\def\wpfigure[#1,#2,#3,#4]{
\begin{figure*}
\vspace*{4mm}
\begin{center}

\includegraphics[width=#4]{#1} 

\vspace*{-3mm}\caption[]{#2
} \label{#3}

\vspace*{-5mm}
\end{center}
\end{figure*}}

\def\wprfigure[#1,#2,#3,#4,#5]{
\begin{figure*}
\vspace*{4mm}
\begin{center}

\includegraphics[width=#4, angle=#5]{#1} 

\vspace*{-3mm}\caption[]{#2
} \label{#3}

\vspace*{-5mm}
\end{center}
\end{figure*}}

\def\DoubleFigureWSlide[#1,#2,#3,#4,#5,#6,#7,#8,#9]{
\begin{figure*}
\vspace*{#9}
\begin{center}
\begin{minipage}{#4}
\includegraphics[width=#4]{#1}
\vspace*{-3mm}\caption{#2
}\label{#3}
\end{minipage}
\hspace{2em}
\begin{minipage}{#8}
\includegraphics[width=#8]{#5}
\vspace*{-3mm}\caption{#6
}\label{#7}
\end{minipage}
\vspace*{-5mm}
\end{center}
\end{figure*}
}

\def\DoubleFigureW[#1,#2,#3,#4,#5,#6,#7,#8]{
\begin{figure*}
\vspace*{0in}
\begin{center}
\begin{minipage}{#4}
\includegraphics[width=#4]{#1}
\vspace*{-3mm}\caption{#2
}\label{#3}
\end{minipage}
\hspace{2em}
\begin{minipage}{#8}
\includegraphics[width=#8]{#5}
\vspace*{-3mm}\caption{#6
}\label{#7}
\end{minipage}
\vspace*{-5mm}
\end{center}
\end{figure*}
}

\def\DoubleFigureWHack[#1,#2,#3,#4,#5,#6,#7,#8]{
\begin{figure*}
\vspace*{0in}
\begin{center}
\begin{minipage}{3in}
\includegraphics[width=#4]{#1}
\vspace*{-3mm}\caption{#2
}\label{#3}
\end{minipage}
\hspace{2em}
\begin{minipage}{3in}
\includegraphics[width=#8]{#5}
\vspace*{-3mm}\caption{#6
}\label{#7}
\end{minipage}
\vspace*{-5mm}
\end{center}
\end{figure*}
}

\def\ddcfigure[#1,#2,#3,#4]{
\begin{figure*}
\vspace*{0.2in}\
\begin{center}
\begin{minipage}[c]{\columnwidth}{
\includegraphics[width=\columnwidth]{#1} 
}\end{minipage}\hspace{0.5in}\
\begin{minipage}[c]{\columnwidth}{
\includegraphics[width=\columnwidth]{#2} 
}\end{minipage} \caption[]{#3}\label{#4}
\end{center}
\end{figure*}
}

\def\ddcfigureSlide[#1,#2,#3,#4,#5]{
\begin{figure*}
\vspace*{#5}\
\begin{center}
\begin{minipage}[c]{3in}{
\includegraphics[height=3in]{#1} 
}\end{minipage}\hspace{0.5in}\
\begin{minipage}[c]{3in}{
\includegraphics[height=3in]{#2} 
}\end{minipage}\vspace*{-0.10in} \caption[]{#3}\label{#4}
\end{center}
\vspace*{-0.4in}\
\end{figure*}
}

\def\cxfigure[#1,#2,#3]{
\begin{figure}
\vspace*{4mm}
\begin{center}
 
\epsfxsize=2.5in\
\epsfbox{#1}\
 
\vspace*{-0.10in}\caption[]{#2
} \label{#3}
 
\vspace*{-5mm}
\end{center}
\vspace*{-2mm}
\end{figure}}

\newcommand{\figWidth}{\columnwidth}

\newcommand{\beforecaption}{\vspace{-.15cm}\begin{spacing}{0.85}}
\newcommand{\aftercaption}{\vspace{-.45cm}\end{spacing}}

\newcommand{\mycaption}[3]{\beforecaption\caption{\label{#1}{\bf \small #2} \em\footnotesize #3}\aftercaption}

\newcommand{\eg}{\textit{e.g.}}
\newcommand{\ie}{\textit{i.e.}}
\newcommand{\etal}{\textit{et al.}}

\newcommand{\KB}{\,KB}

\newcommand{\TB}{\,TB}

\newcommand{\mus}{\mbox{\,$\mu s$}}

\newcommand{\rdma}{RDMA}

\newcommand{\cloverds}{CloverDS}
\newcommand{\cloverkv}{CloverKV}

\newcommand{\cns}{{\texttt{c\&s}}}

\newcommand{\nil}{{\texttt{NULL}}}
\newcommand{\client}{ClvClient}
\newcommand{\control}{ClvCtrl}
\newcommand{\repbit}{$B_r$}
\newcommand{\togclist}{ToGCList}
\newcommand{\epochlist}{OvflowList}
\newcommand{\freelist}{FreeList}
\newcommand{\captbl}{CapTbl}
\newcommand{\readabortT}{$T_r$}
\newcommand{\epochT}{$T_e$}

\newcommand{\nvm}{PM}
\newcommand{\dpm}{DPM}
\newcommand{\cn}{CN}
\newcommand{\metaserver}{MS}
\newcommand{\sdirect}{DPM-Direct}
\newcommand{\scentral}{DPM-Central}
\newcommand{\ssep}{DPM-Sep}
\newcommand{\dir}{DirectDS}
\newcommand{\dircrc}{DirectDS-C}
\newcommand{\dirlock}{DirectDS}
\newcommand{\ctrlredo}{CentralDS}
\newcommand{\sepredo}{SepDS}
\newcommand{\coord}{coordinator}
\newcommand{\SY}{}
\newif\ifremark
\long\def\remark#1{
\ifremark%
        \begingroup%
        \dimen0=\columnwidth
        \advance\dimen0 by -1in%
        \setbox0=\hbox{\parbox[b]{\dimen0}{\protect\em #1}}
        \dimen1=\ht0\advance\dimen1 by 2pt%
        \dimen2=\dp0\advance\dimen2 by 2pt%
        \vskip 0.25pt%
        \hbox to \columnwidth{%
                \vrule height\dimen1 width 3pt depth\dimen2%
                \hss\copy0\hss%
                \vrule height\dimen1 width 3pt depth\dimen2%
        }%
        \endgroup%
\fi}
\remarktrue

\begin{document}

\pagestyle{plain}




\newcommand{\mm}{mm$^2$}
\newcommand{\figtitle}[1]{\textbf{#1}}
\newcommand{\us}{$\mu$s}
\newcommand{\fixme}[1]{{\color{red}\textbf{#1}}}

\definecolor{pink}{rgb}{1.0,0.47,0.6}
\newcommand{\adrian}[1]{{\color{green}\textbf{#1}}}
\newcommand{\laura}[1]{{\color{pink}\textbf{#1}}}
\newcommand{\shinyeh}[1]{{\color{red}\textbf{SY: #1}}}
\newcommand{\shinyehEdit}[2]{{\color{red} \sout{#1} ==> \textbf{#2}}}
\newcommand{\shinyehRM}[1]{{\color{red}\textbf{SY-REMOVE: }}{\color{gray}\textbf{\sout{#1}}}}
\newcommand{\yiying}[1]{{\color{blue}\textbf{#1}}}
\newcommand{\arup}[1]{{\color{yellow}\textbf{#1}}}
\newcommand{\hungwei}[1]{{\color{purple}\textbf{#1}}}
\newcommand{\yizhou}[1]  {\noindent{\color{blue} {\bf \fbox{Yizhou}     {\it#1}}}}

\newcommand{\note}[2]{\fixme{$\ll$ #1 $\gg$ #2}}


\twocolumn[
\begin{@twocolumnfalse}
\begin{center}
{\large\bf Building Atomic, Crash-Consistent Data Stores with Disaggregated Persistent Memory}
\end{center}
\centerline{Shin-Yeh Tsai, Yiying Zhang}
\centerline{\em Purdue University}
\smallskip

\end{@twocolumnfalse}
]

\thispagestyle{plain}

\section*{Abstract}

Byte-addressable persistent memories ({\em \nvm}) has finally made their way into production.
An important and pressing problem that follows is how to deploy them in existing datacenters. 
One viable approach is to attach \nvm\ as self-contained devices to the network as {\em disaggregated persistent memory}, or {\em \dpm}.
\dpm\ requires no changes to existing servers in datacenters;
without the need to include a processor, \dpm\ devices are cheap to build;
and by sharing \dpm\ across compute servers, they offer great elasticity and efficient resource packing.

This paper explores different ways to organize \dpm\ and to build data stores with \dpm.
Specifically, we propose three architectures of \dpm:
1) compute nodes directly access \dpm\ ({\em \sdirect});
2) compute nodes send requests to a coordinator server, which then accesses \dpm\ to complete a request ({\em \scentral});
and 3) compute nodes directly access \dpm\ for data operations and communicate with a global metadata server for the control plane ({\em \ssep}). 
Based on these architectures, we built three atomic, crash-consistent data stores.
We evaluated their performance, scalability, and CPU cost with micro-benchmarks and YCSB.
Our evaluation results show that \sdirect\ has great small-size read but poor write performance;
\scentral\ has the best write performance when the scale of the cluster is small but performs poorly when the scale increases;
and \ssep\ performs well overall. 
\section{Introduction}
\label{sec:introduction}

After year's of research, engineering, and commercializing efforts, persistent memory ({\em \nvm}),
non-volatile memories that can be attached to the main memory bus, 
is finally coming to market~\cite{Intel-Optane, Intel-Optane-News}.
As promised, \nvm\ can be accessed like memory and it offers persistence, high density, 
and performance that is orders of magnitude faster than flash.
It has the potential to significantly improve the efficiency and reduce the cost of large-scale data-intensive applications.
An immediate question that follows is how to utilize \nvm\ and deploy it in existing datacenters.

We believe that a promising approach is to directly attach \nvm\ to the network to form
{\em disaggregated persistent memory}, or {\em \dpm}.
A \dpm\ device only needs a network interface, a hardware \nvm\ controller, and some \nvm;
it requires no server packaging or any processors.
Datacenters owners can use normal servers as compute nodes ({\em \cn{}}s) and store data in \dpm.

The \dpm\ model offers several key benefits.
First, unlike the alternative approach of attaching \nvm\ to a server,
\dpm{}s can be integrated into current datacenters without any disruption to existing servers.
Second, without the need for a processor or a server to host \dpm, 
the monetary and energy cost of \dpm\ is low.
Third, multiple \cn{}s can share one \dpm\ device and one \cn\ can store data on multiple \dpm{}s.
Doing so enables better resource packing than attaching and confining the usage of \nvm\ to a single node~\cite{Greenberg08-SIGCOMM,Gu17-NSDI,Barroso-COMPUTER,Quasar-ASPLOS,PowerNap}.
Fourth, the \dpm\ model offers great elasticity,
since \dpm{}s can be freely added, removed, and replaced. 
The amount of \cn{}s and \dpm{}s can scale {\em independently}.
Finally, although accessing \dpm{}s involves network communication,
this cost is becoming lower as datacenter network speed improves quickly~\cite{Mellanox-Switch,Mellanox200Gbps}. 

Despite its benefits, the \dpm\ model presents new challenges.
Without any processing power, accesses to \dpm{}s 
have to come all from the network, or {\em one-way}.
\dpm{}s cannot perform any management tasks of its own memory resources.
Finally, each \dpm\ can fail independently from \cn{}s or other \dpm{}s.
These unique issues of \dpm{}s have not been addressed in traditional 
distributed storage or distributed memory systems.

To confront the challenges and to explore the design tradeoffs of the \dpm\ model,
we propose three architectures of organizing \dpm{}s (Figure~\ref{fig-arch}(b) to \ref{fig-arch}(d)).
The first architecture, {\em \sdirect}, lets \cn{}s directly access \dpm{}s with one-way operations.
This architecture is cheap to build.
With a fully-distributed architecture, \sdirect\ avoids any throughput bottleneck.
However, it is hard and costly to synchronize concurrent accesses to \dpm{}s from multiple \cn{}s~\cite{OSDI-RPC}.
It is also difficult for \cn{}s to manage memory resources in \dpm{}s.

The second architecture, {\em \scentral}, uses a central server (the {\em \coord}) to orchestrate the accesses from \cn{}s to \dpm{}s
and to manage \dpm\ resources.
\cn{}s can talk to the \coord\ with two-way communication (\eg, through RPC);
the \coord\ orchestrates concurrent \cn\ requests and issues one-way accesses to \dpm{}s.
The \coord\ also stores all metadata and performs metadata operations locally.
Having a central point of the \coord\ makes it easy to manage \dpm{}s and to coordinate concurrent requests,
but the \coord\ can become the performance bottleneck.

To remedy the performance and scalability limitations of \sdirect\ and \scentral, 
we propose a third architecture, {\em \ssep}.
The main idea of \ssep\ is to separate the data plane from the \SY{control plane}.
On the data plane, \cn{}s directly access \dpm{}s.
On the control plane, we use a metadata server ({\em \metaserver}) to handle
all metadata operations and manage \dpm\ resources.
\cn{}s talk to the \metaserver\ with two-way communication to fetch or update metadata.
The \metaserver\ makes it easy and more efficient to perform control plane tasks,
while not being the performance bottleneck on the data path.

{
\begin{figure*}
\begin{center}
\centerline{\includegraphics[width=0.9\textwidth]{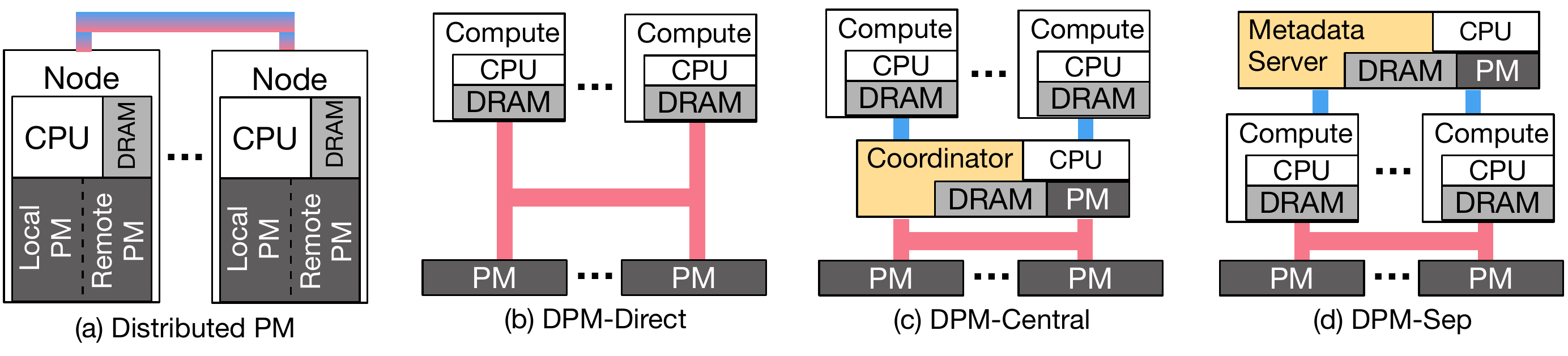}}
\vspace{-0.15in}
\mycaption{fig-arch}{\nvm\ Organization Comparison.}
{
Blue bars indicate two-way communication and pink ones indicate one-way communication.
Bars with both blue and pink mean support for both.
}
\end{center}
\vspace{-0.1in}
\end{figure*}
}

Based on these three architectures, we designed three atomic, crash-consistent \dpm\ data store systems. 
All these systems provide the same guarantees that when writing to a data entry, the data entry either has 
all new data (if the write is successfully committed) or all old data (if the write fails), 
and that \cn{}s only read committed data (\ie, the read-committed isolation level).
These properties hold even when a \dpm\ crashes during the write and recovers afterwards.

On top of the \sdirect\ architecture, we built a data store, {\em \dir}.
To best fit \sdirect\ and provide good performance, we designed \dir\ with two principles: 
reducing network round trips (RTTs) between \cn{}s and \dpm{}s
and avoiding frequent \dpm\ management tasks or metadata modifications.
\dir\ uses two spaces for each data entry, one to write uncommitted new data
and one to store committed data. 
Doing so avoid the need for space allocation after a data entry is created. 
\dir\ protects each data entry with a lock stored in \dpm\ and accessed with one-sided RDMA operations.
We employ techniques like error-detecting code to further reduce RTTs.

On top of \scentral, we built {\em \ctrlredo}.
\ctrlredo\ leverages the centralized \coord\ to perform space management, to store metadata,
and to serve as the serializing point for concurrent accesses. 
\cn{}s send RPC read/write requests to the \coord, which uses local locks
to protect concurrent accesses, reads/writes data to \dpm, and updates metadata locally.

On top of \ssep, we built {\em \sepredo}.
On the data path, \sepredo\ performs out-of-place writes that are similar to log-structured writes.
We use a novel data structure that enables \cn{}s to efficiently locate the latest data entry
without the need to communicate with the \metaserver.
For the control path, the \metaserver\ stores all metadata and \cn{}s caches hot metadata.
We move all metadata operations off performance critical path.
To minimize the need for \cn{}s to communicate with \metaserver,
\cn{}s perform lazy, asynchronous, batched reclamation of old data entries.
We also \SY{{\em completely} eliminate} the need for the \metaserver\ to communicate with \dpm{}s;
it manage \dpm{} space {\em without} accessing them.

To sustain non-transient \dpm\ failures, it is not enough to store data in just one \dpm.
For each of the three systems, we added the support of replication on top of our single-copy designs. 
We also utilize the data redundancy to provide better load balancing for reads ---
we dynamically choose which \dpm{} to replicate data to based on the loads of each \dpm. 

We evaluated the three \dpm\ data stores using real servers as \cn{}s, the \coord, the \metaserver, and \dpm\ devices,
all connected with RDMA.
We emulate \nvm\ using DRAM on real machines;
we perform RDMA read to ensure that data is written to the \nvm\ in \dpm{}s~\cite{SNIA-RDMA-PMEM}.
We perform a systematic, extensive set of experiments
to evaluate the latency, throughput, scalability, network traffic, and CPU utilization
of the three \dpm\ data stores using microbenchmarks and YCSB workloads~\cite{Cooper10-CloudCom,YCSB-C}.
Our evaluation results not only confirm findings that are easy to deduce from system designs
(\eg, that \ctrlredo\ scales poorly with \cn{}s and \dpm{}s and that 
the performance of \sepredo\ is overall the best),
but also reveal more subtle findings 
(\eg, that \dir\ only scales well when there is no contention of concurrent accesses
and that \sepredo's good performance rely on \cn{}s being able to cache hot metadata).
Based on our findings, we summarize the tradeoffs of the three \dpm\ data stores 
in Table~\ref{tbl-system-cmp}. 

This paper makes the following contributions.

\begin{itemize}

\item We propose and compare three \dpm\ architectures.

\item We built three \dpm\ data stores. As far as we know, these are the first set of 
publicly-described and publicly-available \dpm\ systems.

\item We provide a detailed design to demonstrate how to best separate 
data plane and control plane under the \dpm\ model.


\item We performed extensive evaluation and learned a set of new findings that can guide future \dpm\ research.

\end{itemize}

The source code of all our \dpm\ systems will be publicly available soon.

\section{Using \nvm\ in Datacenters}
\label{sec:motivation}


Non-volatile memory technologies such as 
3DXpoint~\cite{Intel3DXpoint}, phase change memory ({\em PCM}),
spin-transfer torque magnetic memories ({\em STTMs}), and the memristor
provide byte addressability, persistence, and latency that is within 
an order of magnitude of 
DRAM~\cite{hosomi2005novel,Lee10-pcmquest,lee2010phase,lee2011fast,pcmdataasheet,qureshi2010morphable,NVMDB,yang2013memristive}.
NVMs can attach directly to the main memory bus and we call such NVMs Persistent Memory or {\em \nvm} in this paper. 
\nvm\ is a disruptive technology poised to radically alter the landscape of memory and storage technologies.
It has attracted extensive amount of research efforts over the past year,
most of which were designed for single-node environments~\cite{Coburn11-ASPLOS,Volos11-ASPLOS,Pelley14-ISCA,Memaripour17-EUROSYS,Kolli16-ASPLOS,Condit09-SOSP,Dragojevic14-FaRM,Dulloor14-EuroSys,Xiaojian11-SC,Ou16-Eurosys}.

Despite of these successful prior research efforts, there are at least two remaining challenges to be solved before \nvm{}s can be readily used in datacenters.
First, in datacenter environments, \nvm{}s should support distributed applications.
When using \nvm{}s to store persistent data, they have to provide high availability and reliability (\ie, sustain node failures).
Unfortunately, there are only limited work in the distributed \nvm\ research space~\cite{Zhang15-Mojim,Shan17-SOCC,Lu17-ATC}.
So far, distributed \nvm\ systems~\cite{Shan17-SOCC,Lu17-ATC} have all taken a
model where each node in a cluster includes some amount of \nvm\ used to store data 
that can be accessed both locally and by other nodes (Figure~\ref{fig-arch}(a)).

Second, it is not clear how to deploy \nvm{}s in {\em existing} datacenters.
The distributed \nvm\ model requires \nvm\ to be integrated into existing servers or 
purchasing new servers to host \nvm.  
Since \nvm{}s attach to the main memory bus, only when existing servers have empty DIMM slots
will they be able to host \nvm.
On the other hand, purchasing new servers just to host \nvm\ can waste other 
resources in the new servers.
Moreover, applications that desire to use \nvm\ can only run on these new servers.

With these challenges, we believe that we should seek new ways to use and deploy
\nvm\ in datacenters that are flexible, cost-effective, reliable, and can perform well.

\section{Disaggregated \nvm}
\label{sec:dpm}


Similar to disaggregated memory~\cite{Lim09-disaggregate,Lim12-HPCA} and other resource disaggregation systems~\cite{Shan18-OSDI,HP-TheMachine,FireBox-FASTKeynote},
disaggregated \nvm\ is an architecture that attaches \nvm\ devices directly to the network
and lets servers ({\em \cn{}s}) access them across the network.
These \nvm\ devices do not have any local processing units and only have a hardware controller and a network interface
(we simply call a disaggregated \nvm\ device a \dpm\ in this paper).
The \dpm\ model organizes \dpm{}s as a pool of \nvm\ resources that can be used by any \cn{}s.
A \cn\ can store data on multiple \dpm{}s and one \dpm\ can host data for multiple \cn{}s.

The \dpm\ model offers a cost-effective way to deploy \nvm\ in datacenters. 
Without any processor or machine packaging, \dpm{}s are cheap to build.
They can easily integrate into existing datacenters without disruption to existing servers.
The \dpm\ model also shares many benefits with other resource disaggregation proposals~\cite{Shan18-OSDI,Gao16-OSDI}:
it offers high resource packing efficiency, since data can be allocated at any \dpm;
datacenters can grow \dpm{}s independent from other servers;
it is easy to add, remove, and upgrade \dpm{}s,
and \dpm{}s can fail independently without affecting other servers.

However, building an efficient \dpm\ data store system is not easy.
A major technical hurdle is the complete lack of computation power at \dpm{}s.
Different from traditional distributed storage and memory systems, 
\dpm{}s can only be accessed {\em and} managed from remote.
It is especially hard to provide good performance with concurrent data accesses.
In addition, \dpm{}s can fail independently and such failures have to 
be handled properly to ensure data reliability and high availability.

{
\begin{table*}[th]\footnotesize
\begin{center}
\begin{tabular}{ l | c | c | c | c | c | c | l}
\footnotesize System & \footnotesize Cost & \footnotesize R-RTT & \footnotesize W-RTT(rep) &\footnotesize Scalability & \footnotesize Data & \footnotesize Metadata & \footnotesize Performance\\
\hline
\dirlock\  & low  	& 3 & 6(6) & w/ \dpm{}$\dagger$    & large	&       large   & OK write performance when no contention\\
\dircrc\   & low 	& 1 & 6(6) & w/ \dpm{}$\dagger$    & large	&       large   & Best for small-sized read, not good otherwise\\
\ctrlredo\ & high  	& 2 & 3(3) & Neither	& small	&       small   & Best for small-scale writes, not good for reads \\
\sepredo\  & medium 	& 1 & 3(4) & w/ both	& small*	&       medium  & Good overall when \cn{}s can cache hot metadata \\

\end{tabular}
\end{center}
\mycaption{tbl-system-cmp}{Comparison of \dpm\ Data Stores.}
{
The Cost column represents energy and monetary cost to build \SY{respective} \dpm\ data stores. 
The R-RTT and W-RTT(rep) columns show the number of RTTs required to perform a read and a write (with replication).
All RTT values are measured when there is no contention.
The Scalability column shows if a system is scalable with the number of \cn{}s, 
the number of \dpm{}s, both, or neither.
$\dagger$ only scalable when there is no contention.
The data and metadata columns show the space needed to store a data entry and its metadata.
* under common scenario where reclamation can keep up with the speed of foreground write.
}
\end{table*}
}
\section{\dpm\ Data Stores}
\label{sec:design}

This section first describes the interface of all our \dpm\ data stores and their common features.
We then present the three data stores, \dir, \ctrlredo, and \sepredo.
Finally, we discuss failure handling and load balancing in these data stores.

\subsection{System Interface and Overview}
\label{sec:systemoverview}
To confront the challenges of \dpm, we propose three architectures of \dpm\
and built three data stores on top of these architectures.
Figure~\ref{fig-dpm-rw} illustrates the read and write operation flow of these systems
and Table~\ref{tbl-system-cmp} summarizes the tradeoffs of these systems.
We will explain Figure~\ref{fig-dpm-rw} and Table~\ref{tbl-system-cmp} \SY{in detail in} \S\ref{sec:direct} to \S\ref{sec:sepplane}.

\ulinepara{Interface and guarantees.}
The current data model that our three data stores support is a simple key-value store, 
but these systems can be extended to other data models.
Users can create, read (get), write (put), and delete a key-value entry.
Different \cn{}s can have shared access to the same data.
We manage the consistency of concurrent data accesses in software
instead of relying on any hardware-provided coherence like~\cite{Genz-citation,ccix-citation,OpenCAPI}. 

All our \dpm\ data stores ensure atomicity of an entry across concurrent readers and writers.
A successful write indicates that the data is committed (atomically),
and reads only see committed value.
We choose \SY{single-key} atomic write and read committed
because these consistency and isolation levels are widely used in many data store systems 
and can be extended to other levels.

Since our \dpm\ systems store persistent data, it is important to provide data reliability and high availability. 
Our \dpm\ systems guarantee the consistency of data when crashes happen.
After restart, each data entry is guaranteed to either only have new data values or old ones.
In addition, all our three systems provide replication across \dpm{}s
to ensure that data is still available even after losing $N-1$ \dpm{}s (when the degree of replication is $N$).

\ulinepara{Network layer.}
We choose RDMA as the network layer that connects all servers and \dpm{}s,
but most of our designs are applicable to other network systems that can perform both one-way and two-way communication.
We use RDMA's RC (Reliable Connection) mode which supports one-sided RDMA operations and ensures lossless and ordered packet delivery.
Similar to prior solutions~\cite{Dragojevic14-FaRM,Tsai17-SOSP}, we solve RDMA's scalability issues 
using memory huge page or physical memory to register memory regions with RDMA NICs.

\ulinepara{Ensuring data persistence.}
For data to be persistent in \dpm, it is not enough to just perform a remote write.
After a remote write (\eg, RDMA write), the data can be in NIC, PCIe hub, or \nvm.
Only when the data is written to \nvm\ can it sustain power failure.
To ensure this data persistence, we follow the guidance of SNIA~\cite{SNIA-RDMA-PMEM} and Mellanox~\cite{RDMA-RFC, IBSPEC}
by performing a remote read to ensure that data is actually in \nvm.
Since we use RDMA RC which guarantees ordered data delivery
and PCIe also follows ordering~\cite{PCIE-Spec}, 
we only read the last byte of a data entry to verify its persistence~\cite{SNIA-RDMA-PMEM}.

\subsection{Direct Connection}
\label{sec:direct}
{
\begin{figure*}
\begin{center}
\centerline{\includegraphics[width=\textwidth]{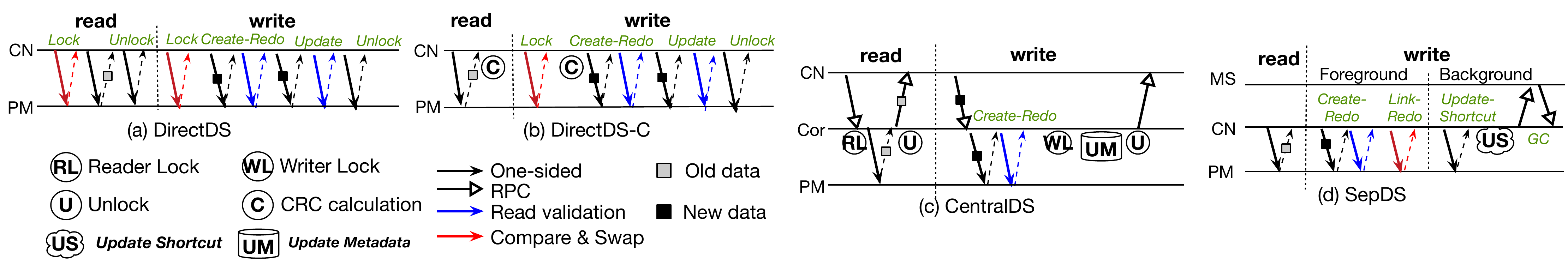}}
\vspace{-0.15in}
\mycaption{fig-dpm-rw}{Read/Write Protocols of \dpm\ Systems.}
{
}
\end{center}
\vspace{-0.2in}
\end{figure*}
}
The \sdirect\ architecture (Figure~\ref{fig-arch}(b)) connects \cn{}s directly to \dpm{}s.
\cn{}s perform un-orchestrated, direct accesses to \dpm{}s using RDMA one-sided operations.
Under \sdirect, performing metadata and control operations from \cn{}s is hard and costly (\eg, by performing distributed coordination across \cn{}s).
Thus, we made two design choices when building \dir. 

First, we use two spaces for each data entry, one to store committed data where reads go to (we call it the {\em committed space}) 
and one to store in-flight, new data ({\em un-committed space}).
Doing so avoids dynamic space allocation and de-allocation.
Second, to avoid reading and writing metadata from \dpm{}s and the cost of ensuring metadata consistency under concurrent accesses,
\cn{}s in \sdirect\ locally store all the metadata of key-value entries,
including the key of a value and the location of its committed and uncommitted spaces.

The only distributed coordination across \cn{}s needed in \sdirect\ is during the creation and deletion of a key-value entry.
We currently use Memcached~\cite{fitzpatrick2004distributed} \SY{as a metadata server} to assist entry creation and delete,
but other distributed consensus systems can also work.

Figure~\ref{fig-dpm-rw}(a) illustrates the read and write protocol of \dir.
\dir\ uses locks to isolate data entries from concurrent read and write accesses. 
Each entry has its own lock and we associate a 8-byte value at the beginning of each data entry to implement its lock.
A \cn\ performs a one-sided RDMA \cns\ (compare-and-swap) operation to the value to acquire the lock
(\eg, comparing whether the value is 0 and if so setting it to 1).
To release the lock, the \cn{} simply performs an RDMA write and sets the value to 0.

Our lock implementation leverages the unique feature of the \dpm\ model that all memory accesses to \dpm{}s come from the network (\ie, the NIC). 
Without processor's accesses to memory, DMA guarantees 
that network atomic operations like \cns\ are atomic~\cite{Tang10-HPCA, Daglis16-MICRO}.
Note that an RDMA \cns\ operation to an in-memory value which can also be accessed locally at the same time 
does not guarantee the atomicity of the value~\cite{Wei15-SOSP, ibverbs,Daglis16-MICRO},
and thus it cannot be used in distributed \nvm\ systems in the same way.

\ulinepara{Read.}
To read a data entry, a \cn\ uses its stored metadata to find the location of the data entry's committed space (and the first 8-byte lock).
It first acquires a lock, then performs an RDMA read, and finally releases the lock.
Locking reads ensures that \cn{}s will not read intermediate value during concurrent writes. 
The read latency is 3 RTTs when there is no contention, with one RTT used for data read.
Under contention, the \cns\ operation would fail and 
\cn{}s will keep retry until succeed.

\ulinepara{Write.}
To write a data entry, a \cn\ first locates the entry and locks it.
Afterwards, the \cn\ writes the new data to the un-committed space.
To sustain crashes, the \cn\ issues an RDMA read to the last byte of the un-committed space to validate that it is actually written to \nvm.
This uncommitted data serves as the {\em redo} copy that will be used during recovery if a crash happens. 
The \cn\ then writes the new data to the committed space with an RDMA write and validates it with an RDMA read.
At the end, the \cn\ releases the lock.
The total write latency is 6 RTTs (without contention), two of which involve data read/write.

\ulinepara{Avoid read lock with CRC.}
\dirlock\ uses lock to ensure the read-committed isolation level at the cost of two RTTs to acquire and release the lock for each read. 
Instead of lock, we can use an error-detecting code for each data entry to detect incomplete data.
{\em \dircrc} (Figure~\ref{fig-dpm-rw}(b)) uses the CRC code for this purpose.

To perform a read, a \cn\ simply issues an RDMA read to fetch the data
and then calculates and validates its CRC.
Thus, the read latency of \dircrc\ is one RTT plus the CRC calculation time.
Writes in \dircrc\ is similar to \dirlock, except that before writing the new data,
the \cn\ needs to first calculate and attach a CRC to the new data entry.

\ulinepara{Discussion.}
As we will see in \S\ref{sec:results},
as expected, \sdirect\ data stores scale well when there is no contention of concurrent accesses to data entries.
More surprising is that they scale very poorly when contention happens, especially with write. 
In general, the write performance is not good because of the high RTTs.
But write performs especially poorly under contention,
because multiple \cn{}s will all try to acquire the lock with the \cns\ operation
and most of them will experience a lot of \cns\ failures.
However, \dircrc\ yields the best read performance when read size is small,
since it only requires one lock-free RTT and it is fast to calculate small CRC.

\sdirect\ systems also require large space for both data and metadata.
For each data entry, it doubles the space because of the need to store two copies of data.
The metadata overhead is also high, since \cn{}s have to store all metadata.

\subsection{Connecting Through Coordinator}
\label{sec:central}

Most limitations of \sdirect\ come from the fact that there is no central coordination of data, metadata, or management operations.
Specifically, \sdirect\ systems have to write data twice, once to the un-committed and once to the committed space,
because \cn{}s in \sdirect\ only know a fixed location to read committed data.
The \scentral\ architecture (Figure~\ref{fig-arch}(c)) takes the opposite design choice
and uses one \coord\ to orchestrate all data accesses and to perform metadata and management operations.
All \cn{}s send RPC requests to the \coord\ (we use the HERD~\cite{Kalia14-RDMAKV, Kalia16-ATC} RPC system for this purpose).
The \coord\ handles RPC requests by performing read/write requests to \dpm{}s.
To improve application throughput, we use multiple threads at the \coord\ to handle RPC requests.

Since all requests go through the \coord,
it can serve as the serialization point for concurrent accesses to a data entry.
We simply use a local read/write lock for each data entry at the \coord\ as the synchronization of multiple \coord\ threads.
In addition to orchestrating data accesses, the \coord\ performs all space allocation and de-allocation of data entries.
The \coord\ uses its local \nvm\ to persistently store all the metadata for a data entry including its key, its location, and a read/write lock.
With the \coord\ handling all read requests, it can freely direct a read to the latest location of committed data.
Thus, it does not need to maintain the same location for committed data and changes the location of committed data after each write.

\ulinepara{Read.}
To perform a read, a \cn\ sends an RPC read request to the \coord.
The \coord\ finds the location of the entry's committed data using its local metadata,
acquires its local lock of the entry,
reads the data from the \dpm\ using an RDMA read,
releases the lock, 
and finally replies to the \cn's RPC request.
The total read latency (from \cn's perspective) is 2 RTTs, both containing data.

\ulinepara{Write.}
After receiving a write RPC request from a \cn, the \coord\ allocates a new space in a \dpm\ for the new data.
It then writes the data and validates it with an RDMA read. 
Note that we do not need to lock (either at \coord\ or at \dpm) during this write, 
since it is an out-of-place write to a location that is not exposed to any other \coord\ RPC handlers.

After successfully verifying the write, the \coord\ updates its local metadata of where the committed version of 
the data entry is and flushes this new metadata to its local \nvm\ for crash recovery 
(by performing CPU cache flushes and memory barrier instructions~\cite{Chen16-MICRO}).
Since concurrent \coord\ RPC handlers can update the same information of where the latest data entry is,
we use a local lock to protect this metadata change.
The total write latency without contention is 3 RTTs, with two of them containing data \SY{and one for validation}.

\ulinepara{Discussion.}
\ctrlredo\ largely reduces write RTTs over \dir\
and thus has good write performance when the scale of the cluster is small.
However, from our experiments, the \coord\ soon becomes the performance bottleneck
when either the number of \cn{}s increases or the number of \dpm{}s increases.
\ctrlredo's read performance is also worse than \dircrc\ with the extra hop between a \cn\ and the \coord.
In addition, the CPU utilization of the \coord\ is high, since it needs to have a high amount of RPC handlers
to sustain parallel requests from \cn{}s (\S\ref{sec:results}).
However, unlike \sdirect, \cn{}s in the \sdirect\ architecture does not need to store any metadata.
{
\begin{figure}[th]
\begin{minipage}{\figWidth}
\begin{center}
\centerline{\includegraphics[width=\figWidth]{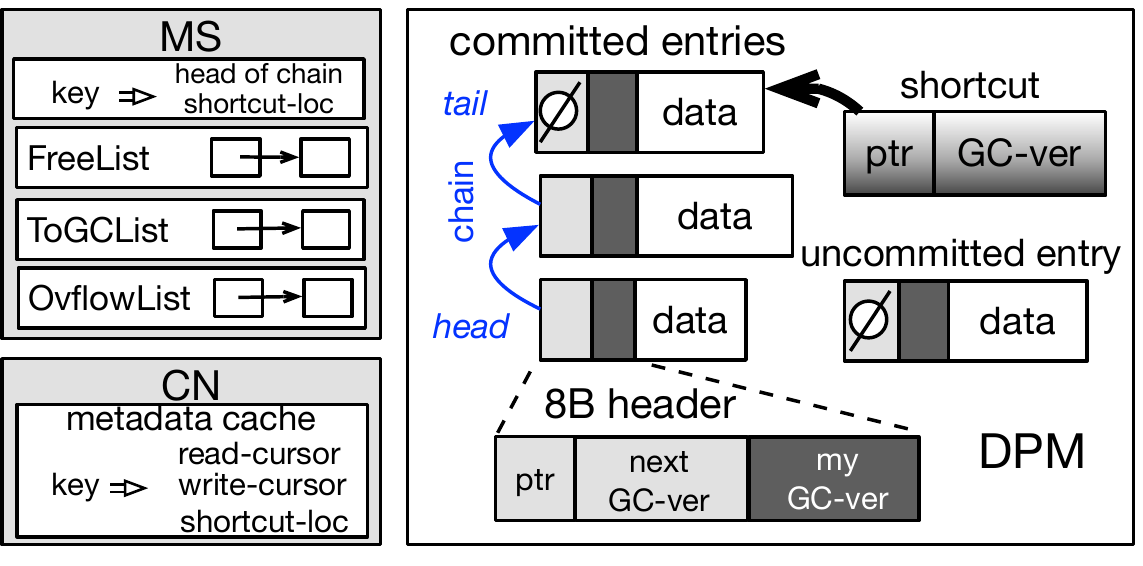}}
\vspace{-0.1in}
\mycaption{fig-sepredo}{\sepredo\ System Design.}
{
}
\end{center}
\end{minipage}
\vspace{-0.1in}
\end{figure}
}
\subsection{Separating Data and Control}
\label{sec:sepplane}

The main issue with \sdirect\ is its poor write performance. 
\ctrlredo\ improves the write performance but suffers from the scalability bottleneck of the central \coord.
To solve these problems of the first two \dpm\ architectures, we propose a third architecture, \ssep\ (Figure~\ref{fig-arch}(d)),
and a data store designed for it, {\em \sepredo}.
The main idea of \ssep\ is to separate the data plane from the control plane.
It lets \cn{}s directly access \dpm{}s for all data operations
and uses a metadata server (\metaserver) for all control plane operations.

The \metaserver\ stores metadata of all data entries in its local \nvm.
We keep the amount of metadata small,
and 1\TB\ of \nvm\ (a conservative estimation of the size of \nvm\ a server can host)
can store metadata for 64\TB\ data at the granularity of 1\KB\ per data entry.
\cn{}s cache metadata of hot data entries;
under memory pressure, \cn{}s will evict metadata according to an eviction policy (we currently support FIFO and LRU).

\sepredo\ aims to deliver scalable, good performance at the data plane
and to avoid the \metaserver\ being the bottleneck at the control plane.
Our overall approaches to achieve these design goals include:
1) moving all metadata operations off performance critical path,
2) using lock-free data structures to increase scalability,
3) employing optimization mechanisms to reduce network round trips for data accesses,
and 4) leveraging the unique atomic data access guarantees of \dpm.
Figure 3 illustrates the data structures used in \sepredo. 

\subsubsection{Data Plane}
\label{sec:sepredodata}
To achieve our data plane design goal, we propose a new mechanism 
to perform lock-free, fast, and scalable reads and writes.
The basic idea is to allow multiple committed versions of a data entry in \dpm{}s
and to link them into a {\em chain}.
Each committed write to a data entry will move its latest version to a new location.
To avoid the need to update \cn{}s with the new location,
we use a self-identifying data structure to let \cn{}s be able to find the latest version.

We include a {\em header} with each version of a data entry, which contains a pointer and some metadata bits used for garbage collection.
The pointers chain all versions of a data entry together in the order that they are written.
A \nil\ pointer indicates that the version is the latest.

A \cn{} acquires the header of the chain head from the \metaserver\ at the first access to a data entry.
It then caches the header locally to avoid the overhead of contacting \metaserver\ on every data access.
As a \cn\ reads or writes an entry, it advances its cached header. 
We call a \cn-cached header a {\em cursor}.

\ulinepara{Read.}
\sepredo\ reads are lock-free.
To read a data entry, the \cn\ performs a {\em chain walk}. 
The chain walk begins with fetching the data entry its current cursor points to.
It then follows the pointer in the following entries
until it reaches the last entry.
All steps in the chain walk use one-sided RDMA reads.
After a chain walk, the \cn\ updates its cursor to the last entry.

A chain walk can be slow with long chains when a cursor is not up to date~\cite{Wu17-VLDB}.
Inspired by skip-list~\cite{Skiplist}, we propose to solve this issue by using a {\em shortcut} to directly point to a newer entry.
The shortcut of a data entry is stored in \dpm\ and the location of the shortcut never changes during the lifetime of the data.
\metaserver\ stores the locations of all shortcuts and \cn{}s cache the hot ones.
Shortcuts are {\em best effort} in that they are intended but not enforced to always point to 
the last version of an entry.

The \cn{} issues a chain walk read and a shortcut read in parallel.
It returns to user when the faster one returns and discards the other result.
Note that we do not replace chain walks completely with shortcut reads,
since shortcuts are updated asynchronously in the background and may not be updated as fast as the cursor.
When the \cn{} has a pointer that points to the latest version of data,
a read only takes 1 RTT.

\ulinepara{Write.}
\sepredo\ never overwrites existing data entries  
and performs a lock-free out-of-place write before linking the new data to an entry chain.
To write a data entry, a \cn\ first selects a free \dpm\ buffer assigned to it by \metaserver\ in advance 
(see \S~\ref{sec:controller}).
It performs a one-sided RDMA write to write the new data to this buffer
and then issues a read of the last byte to ensure that the data is written in \nvm.
Afterwards, the \cn\ performs an RDMA \cns\ operation to link this new entry to the tail of the entry chain.
Specifically, the \cns\ operation is on the header that \cn's cursor points to.
It compares if the pointer in the header is \nil\ and swaps the pointer to point to the new entry.
If the \cns\ succeeds, we treat this data as {\em committed} and return the write request to the user. 
If the pointer is not \nil, it means that the cursor does not point to the tail of the chain
and we will do a chain walk to reach the tail and then do another \cns.

Afterwards, the \cn\ uses a one-sided RDMA write to update the shortcut of the entry 
to point to the new data entry. This step is off the performance critical path.
The \cn\ also updates its cursor to the newly written data entry. 
We do not invalidate or update other \cn{}s' cursors at this time
to improve the scalability and performance of \sepredo.

\sepredo' chained structure and write mechanism ensure that writers do not block readers and readers do not block writers.
They also ensure that readers can only view committed data.
Without high write contention to the same data entry, one write takes only 3 RTTs.

\ulinepara{Retire.}
After committing a write, a \cn\ can {\em retire} the old data entry, 
indicating that the entry space can be reclaimed.
To improve performance and minimize the need to communicate with the \metaserver,
\cn{}s perform lazy, asynchronous, batched retirement of old data entries in the background.
We further avoid the need for \metaserver\ to invalidate \cn{}-cached metadata
using a combination of timeout and epoch-based garbage collection.

\subsubsection{Control Plane}
\label{sec:controller}
\cn{}s communicate with the \metaserver\ using two-sided operations for all metadata operations.
The \metaserver\ performs all types of management of \dpm{}s. 
It manages physical memory space of \dpm,
stores the location and shortcut of a data entry.
We carefully designed these \metaserver\ functionalities to achieve good performance and scalability.

\ulinepara{Space allocation.}
With the data plane out-of-place write model, \sepredo\ has high demand for \dpm\ space allocation.
We use an efficient space allocation mechanism
where \metaserver\ packages free space of all \dpm{}s into chunks. 
Each chunk hosts the same size of data entries and different chunks can have different data sizes, 
similar to FaRM~\cite{Dragojevic14-FaRM} and Hoard~\cite{Berger00-ASPLOS}.
Instead of asking for a new free entry before every write, 
each \cn\ requests multiple entries at a time from the \metaserver\ in the background.
This approach moves space allocation 
off the critical path of writes and is important to deliver good write performance. 

\ulinepara{Garbage collection.}
\sepredo' append-only chained data structure makes its writes very fast.
But like all other append-only or log-structured data stores, \sepredo\ needs to garbage collect (GC) old data.
We designed a new efficient GC mechanism that does not involve any data movement or communication to \dpm\
and minimizes the communication between \metaserver\ and \cn{}s.

The basic flow of GC is simple: 
the \metaserver\ keeps busy checking and processing incoming retire requests from \cn{}s.
The \metaserver\ decides when a data entry can be reclaimed and puts a reclaimed entry to 
a free list ({\em \freelist}).
It gets free entries from this list when \cn{}s request for more free buffers.
A reclaimed entry can be used by any \cn\ for any new entry,
as long as the size fits.

Although the above strawman GC implementation is simple, 
making GC work correctly, efficiently, and scale well is challenging.
First, to achieve good GC performance, 
we avoid the invalidations of \cn\ cached cursors after reclaiming entries so as to minimize the network traffic between the \metaserver\ and \cn{}s.
However, with the strawman GC implementation, \cn{}s' outdated cursors can cause failed chain walks.
We solve this problem using two techniques:
1), the \metaserver\ does not clear the header (or the content) of a data entry after reclaiming it,
and 2), we assign a {\em GC version} to each data entry.
The \metaserver\ increases the GC version number after reclaiming a data entry.
It gives this new GC version together with the location of the entry 
when assigning the entry as a new free buffer to a \cn, $A$.
Before \cn\ $A$ uses the entry for its new write, 
the entry content at the \dpm\ still has old header and data (with old GC version). 
Other \cn{}s that have cached cursors to this entry 
can thus still use the old pointer to perform chain walk.
\cn{}s differentiate if an entry is its intended data or has already been reclaimed and reused for other data
by comparing the GC version in its cached cursor and the one it reads from the \dpm.
After \cn\ $A$ writes the new data with the new GC version number, 
other \cn{}s that have the old cursors will have a mismatched GC version 
and discard the entry and invalidates their cursors.
Doing so not only avoids the need for \metaserver\ to invalidate cursor caches on \cn{}s,
but also eliminates the need for \metaserver\ to access \dpm{}s during GC.

The next challenge is related to our targeted guarantee of read isolation and atomicity 
(\ie, readers should always read the data that is consistent to its metadata header).
An inconsistent read can happen if the read to a data entry takes long
and during the reading time, this entry has been reclaimed and used to write a new data entry.
We use a read timeout scheme similar to ~\cite{Dragojevic14-FaRM}.
\cn{}s abort a read operation after \readabortT, an agreed value among \cn{}s and the \metaserver. 
The \metaserver\ delays the actual reclamation of an entry to only \readabortT\ time after 
it receives the retire request of the entry.
Specifically, the \metaserver\ leaves the entry in a {\em \togclist} for \readabortT\ 
and then moves it to the \freelist.

The final challenge is the overflow of GC version numbers. 
We can only use limited number of bits for GC version in the header of a data entry (currently 8 bits), 
since the header needs to be smaller than the size of an atomic RDMA operation.
When the GC version of an entry increases beyond the maximum value, 
we will have to restart it from zero.
With just the GC version number and our GC mechanism so far, 
\cn{}s will have no way to tell if an entry matches its cached cursor version
or has advanced by $2^8=256$ versions. 
To solve this rare issue without invalidation traffic to \cn{}s,
we use an epoch-based timeout mechanism.
When the \metaserver\ finds the GC version number of a data entry overflows,
it puts the reclaimed entry into {\em \epochlist}
and waits for \epochT\ time before moving it to the {\em \freelist} that can be assigned to \cn{}s.
All \cn{}s invalidate their own cursors after 
an inactive period of \epochT\ (if during this time, the \cn\ access the entity, it would have advanced the cursor already).
To synchronize epoch time, the \metaserver\ sends a message to \cn{}s after \epochT,
and the \metaserver\ can choose the value of \epochT. 
Epoch message is the only communication the \metaserver\ issues to \cn{}s 
during GC.

\subsubsection{Discussion.}
The \sepredo\ design offers four benefits.
First, \sepredo\ reads and writes are fast, 
with 1 RTT and 3 RTTs respectively when there is no contention.
Even under contention, \sepredo\ still outperforms \dir\ and \ctrlredo.
Achieving this low latency and guaranteeing atomic write and read committed is not easy
and is achieved by the combination of four approaches:
1) ensuring the data path does not involve the \metaserver,
2) reducing metadata communication to the \metaserver\ and moving it off performance critical path,
3) ensuring no memory copy in the whole data path,
and 4) leveraging the unique advantages of \dpm\ to perform RDMA atomic operations.

Second, \sepredo\ scales well with the number of \cn{}s and \dpm{}s,
since its reads and writes are both lock free.
Readers do not block writers or other readers and writers do not block readers. 
Concurrent writers to the same entity only contend for the short period of 
RDMA \cns\ operation.
\sepredo\ also minimizes the network traffic to \metaserver\ 
and the processing load on \metaserver\ to make \metaserver\ scale well with number of \cn{}s and data operations.

Third, we avoid {\em all} data movement or communication between the \metaserver\ and \dpm{}s during GC.
To scale and support many \cn{}s with few \metaserver{}s, we avoid \cn\ invalidation messages completely.
The \metaserver\ does not need to proactively send any other messages to \cn{}s either.
Essentially, the \metaserver\ never {\em pushes} any messages to \cn{}s.
Rather, \cn{}s {\em pull} information from the \metaserver. 

Finally, the \sepredo\ data structure is flexible and can support load balancing very well. 
Different entries of a data entity do not need to be on the same \dpm\ device. 
As we will see in \S\ref{sec:replication} and \S\ref{sec:loadbalancing}, 
this flexible placement is the key to \sepredo' load balancing and data replication needs.

However, \sepredo\ also has its own limitation. 
It requires \cn{}s to cache metadata. 
As we will see in \S\ref{sec:results}, when \cn's local metadata cache becomes small, \sepredo's performance drops.
Thus, \sepredo\ works the best when \cn{}s have enough memory or when data accesses have good temporal locality.
\subsection{Failure Handling}
\label{sec:replication}

\dpm{}s can fail independently from \cn{}s.
A \dpm\ system needs to handle both the transient failure of a \dpm\ (which can be rebooted)
and a permanent failure of one.
For the former, our three \dpm\ systems guarantee crash consistency, 
\ie, after reboot, the \dpm\ can recover all its committed data.
For the latter, we add the support for data replication across multiple \dpm{}s 
to all the three data store systems.
In addition, \ctrlredo\ and \sepredo\ also need to handle the failure of the \coord\ and the \metaserver.

\subsubsection{Recovery from Transient Failures}
We now present how each system recovers from a single \dpm's failure when it restarts. 
We assume that the rest of the system (\eg, \cn{}s, the \coord, the \metaserver) keeps alive.
We will discuss the reliability of the \coord\ and the \metaserver\ in \S\ref{sec:replication}.

\ulinepara{\dir.}
When recovering a \dpm\ in \dir, we need to decide whether to use the data in the 
committed space or the un-committed space (\ie, where the redo copy is).
Note that a crash can happen when writing to the committed space, leaving it in an intermediate state, 
in which case a correct recovery should use the un-committed space (the redo copy).
We use a technique that leverages RDMA's ordered writes in increasing address order~\cite{Taleb18-ATC, Dragojevic14-FaRM}
to ensure the integrity of a data space.
Specifically, \dir\ extends its write data by attaching a unique 8-byte value to the beginning and the end of a data entry,
and writes the extended data entry during its write protocol.
The unique value can be calculated by maintaining a monotonically increasing number at each \cn.
During recovery, we compare the first and last 8 bytes of the committed space. 
A match indicates the committed space has the complete data.
Otherwise, we check the un-committed space and use the same way to tell if it has the complete data.

\dircrc\ does not need this extended write mechanism and can simply validate the data in the committed space with its CRC. 
If the CRC is incorrect, we copy the data from the redo copy to the committed space.

\ulinepara{\ctrlredo.}
Handling the failure of a \dpm\ in \ctrlredo\ is simple, as long as the \coord\ stays alive.
Since \ctrlredo\ performs out-of-place writes and the \coord\ stores the state of all writes,
we can simply use the information in the \coord\ to know what writes have written their redo copies 
but haven't committed yet and what writes have not written redo copies.
For the former case, we advance to the redo copy, and for the latter, we use the original version.


\ulinepara{\sepredo.}
\sepredo' recovery mechanism is also simple.
If a \dpm\ fails before a \cn\ successfully links the new data it writes to the chain (indicating an un-committed write),
the \cn\ simply unsets lock bits \SY{(within a pointer)} of the data entry (releasing the held lock) and discards the new write (by treating the space as unused). 

\subsubsection{Adding Redundancy}
\label{sec:replication}
We now present how we add redundancy to \dpm\ in all the three systems
and how we handle \coord\ and \metaserver\ failures.
With the user-specified degree of replication being $N$,
our data store systems guarantee that data is still accessible after $N-1$ \dpm{}s have failed.

\noindent{\underline{\textit{\textbf{\dir\ and \dircrc.}}}}
In order to sustain \dpm\ failure during a write, we need to replicate both the first write to the un-committed space (the redo copy)
and the second write to the committed space. 
After getting the lock, a \cn\ sends the new data to the un-committed space on $N$ \dpm{}s in parallel.
Afterwards, it performs $N$ read validation, also in parallel. 
Once read validation of all the copies succeeds, the \cn\ writes the data to the committed space of the $N$ \dpm{}s in parallel
and performs a parallel read validation afterwards.

\ulinepara{\ctrlredo.}
To handle a replicated write RPC request, the \coord\ writes multiple copies of the data to $N$ \dpm{}s in parallel
and performs a parallel read validation of them.
After the read validation, the \coord\ updates its metadata to record the new locations of all these copies. 

\ulinepara{\sepredo.}
{
\begin{figure*}[th]
\begin{minipage}{0.7\columnwidth}
\begin{center}
\centerline{\includegraphics[width=1.0\columnwidth]{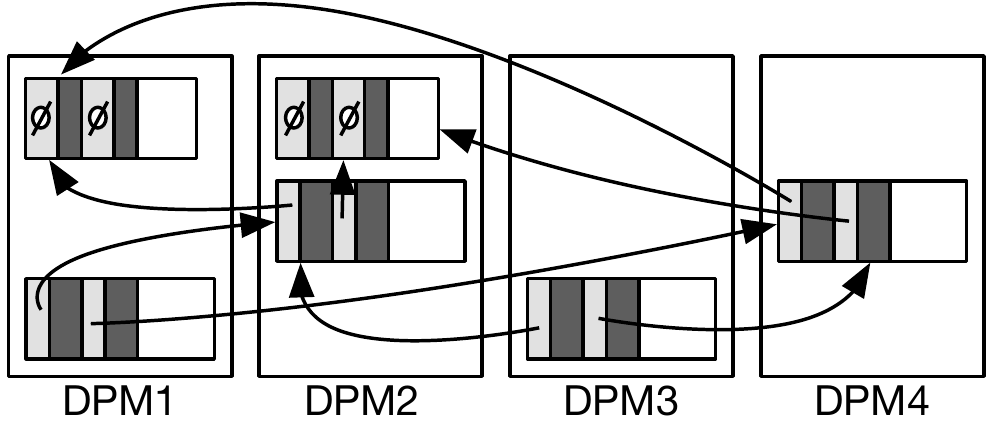}}
\vspace{-0.1in}
\mycaption{fig-rep}{Replicated Data Entity.}
{
A replicated data entity on four \dpm{}s. 
The replication factor is two.
}
\end{center}
\end{minipage}
\vspace{-0.1in}
\begin{minipage}{0.64\columnwidth}
\begin{center}
\centerline{\includegraphics[width=1.0\columnwidth]{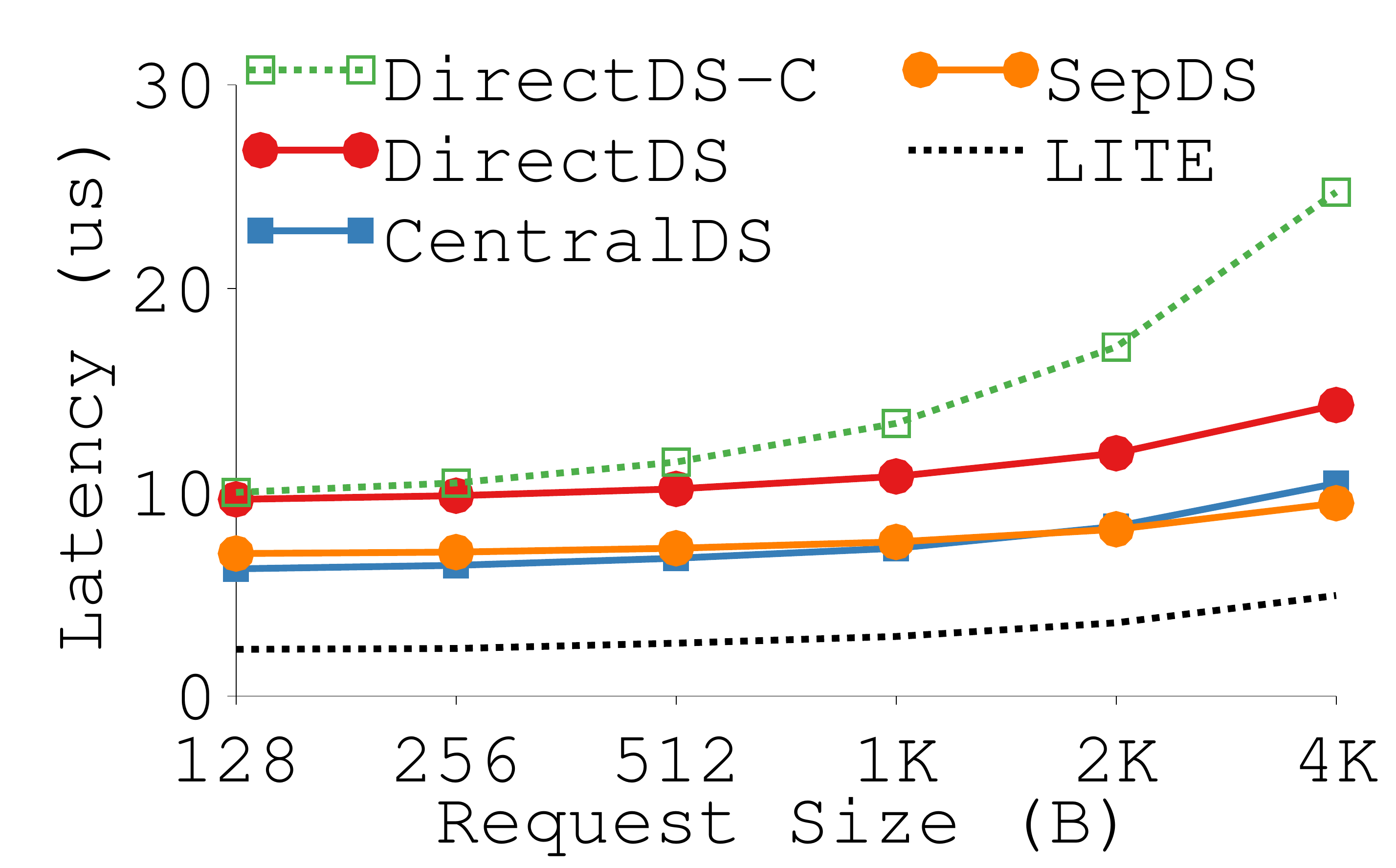}}
\vspace{-0.1in}
\mycaption{fig-dpm-write-latency}{Write Latency}
{
}
\end{center}
\end{minipage}
\begin{minipage}{0.1in}
\hspace{0.01in}
\end{minipage}
\begin{minipage}{0.64\columnwidth}
\begin{center}
\centerline{\includegraphics[width=1.0\columnwidth]{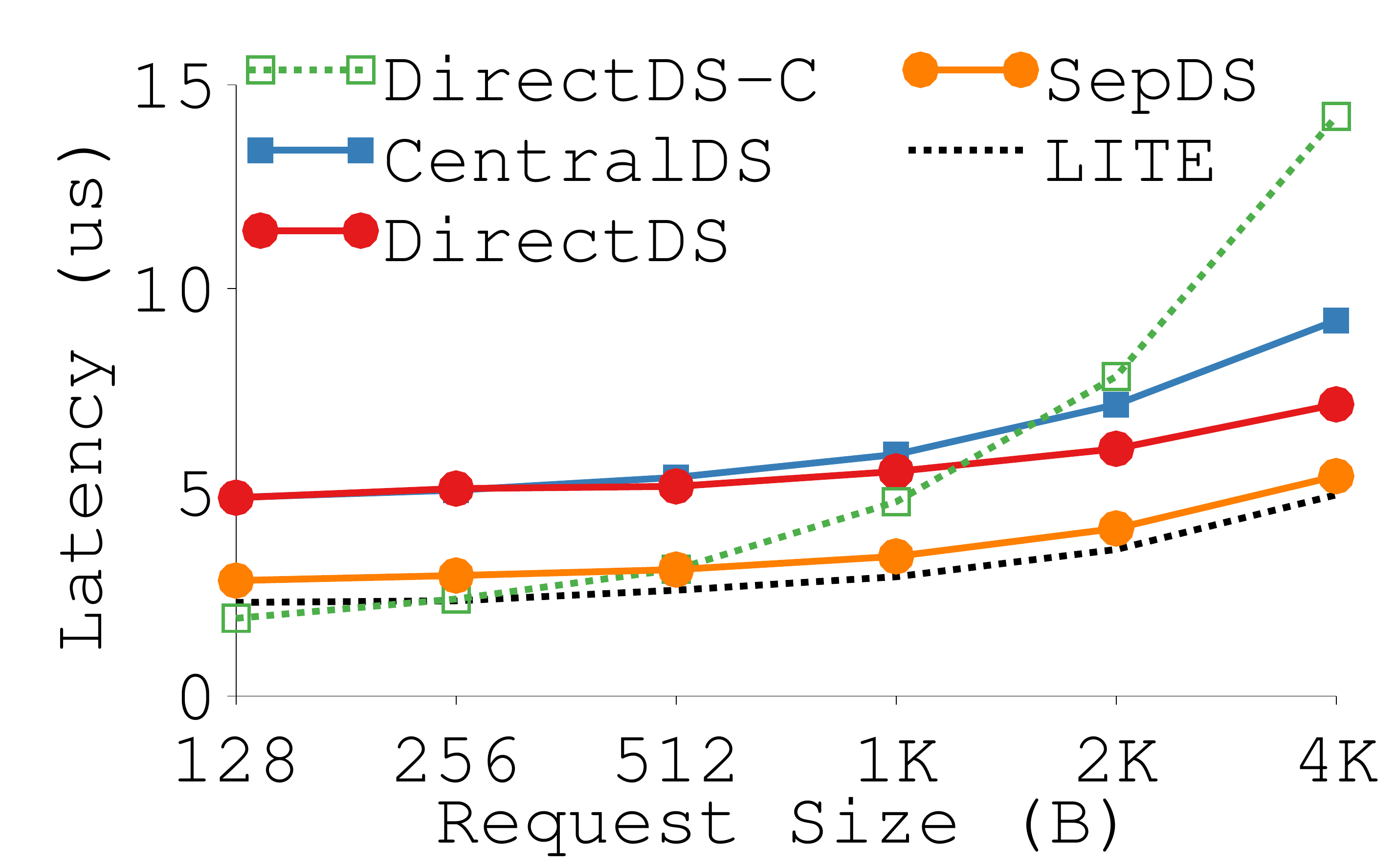}}
\vspace{-0.1in}
\mycaption{fig-dpm-read-latency}{Read Latency}
{
}
\end{center}
\end{minipage}
\end{figure*}
}
We propose a new atomic replication mechanism designed for the \sepredo\ data structure.
The basic idea is to link each data entry version $D_N$ to all the replicas of the next version (\eg, $D_{N+1}^a$, $D_{N+1}^b$, $D_{N+1}^c$ for three replicas)
by placing pointers to all these replicas in the header of $D_N$. 
Figure~\ref{fig-rep} shows an example of replicated data entry.
With this all-way chaining, \sepredo\ can always construct a valid chain
as long as one copy of each version in an entry survives.

Each data entry has a primary copy and one or more secondary copies.
To write a data entry $D_{N+1}$ with $R$ replicas to an entry whose current tail is $D_N$, 
a \cn\ first writes all copies of $D_{N+1}$ to $R$ \dpm{}s.
In parallel, a \cn\ performs a one-sided \cns\ to a bit, $B_w$, in the header of the primary copy of $D_N$
to test if the entry is already in the middle of a replicated write.
If not, the bit will be set, indicating that the entry is now under replicated write.
All the writes and the \cns\ operation are sent out together to minimize latency.

After the \cn\ receives the hardware acknowledgment of all the operations, 
it constructs a header that contains $R$ pointers to the copies of $D_{N+1}$
and writes it to all the copies of $D_N$.
Once the new header is written to all copies of $D_N$, 
the system can recover $D_{N+1}$ from crashes (up to $R-1$ concurrent \dpm\ failure).

\noindent{\underline{\textit{\textbf{Backup \coord\ and \metaserver.}}}}
To avoid the \coord\ or the \metaserver\ being the single point of failure in \ctrlredo\ and \sepredo,
we implement a mechanism to enabling one or more {\em backup} \coord\ (\metaserver),
by having the primary \coord\ (\metaserver) replicate the metadata that cannot be reconstructed (\ie, keys and locations of values)
to the backup \coord\ (\metaserver) when changing these metadata.

\subsection{Load Balancing}
\label{sec:loadbalancing}
With the \dpm\ model, a system will have a pool of \dpm{}s.
Thus, it is beneficial to balance the load to each of them.

With a centralized place to initiate all requests, it is easy for \ctrlredo\ to perform load balancing.
The \coord\ simply records the load to each \dpm\ and directs new writes to the \dpm\ with lighter load.
When \dpm\ is replicated, the \coord\ can also balance read loads by selecting the replica that is on the \dpm\ with lighter load.

We use a novel two-level approach to balance loads in \sepredo:
globally at \metaserver\ and locally at each \cn.
Our global management leverages two features in \sepredo:
1) \metaserver\ assigns all new space to \cn{}s;
and 2) data entries of the same entity in \sepredo\ can be on different \dpm{}s.
To reduce the load on a \dpm, \metaserver\ directs all new writes
to other devices.
At a local level, each \cn\ internally balances the load to different \dpm{}s.
Each \cn\ keeps one bucket per \dpm\ to store free entries. 
It chooses buckets from different buckets for new writes according to its own 
load balancing needs. 

However, balancing loads with the \sdirect\ architecture is hard,
since there is no coordination across \cn{}s.
\section{Evaluation Results}
\label{sec:results}

This section presents the evaluation results of different \dpm\ systems including 
\dirlock, \dircrc, \ctrlredo, and \sepredo.
All our experiments were carried out in a cluster of 14 machines, 
connected with a 40 Gbps Mellanox InfiniBand Switch.
Each machine is equipped with two Intel Xeon E5-2620 2.40GHz CPUs, 128 GB DRAM,
and one 40 Gbps Mellanox ConnectX-3 NIC.


\subsection{Micro-benchmark Results}
We then evaluate \dpm\ data stores' read and write performance
and \SY{compare them} to LITE~\cite{Tsai17-SOSP}. 
We chose LITE for comparison since it offers low latency and 
uses a similar physical memory registration method as our data stores.

Figure~\ref{fig-dpm-write-latency} plots the average write latency with different request size.
LITE performs a write without read validation and only models the latency of un-validated writes.
Its latency is thus the lowest.
Among \dpm\ systems, \sepredo\ and \ctrlredo\ achieve the best write latency.
\sepredo\ outperforms \ctrlredo\ slightly when request size is big
because of its smaller network traffic.
\dirlock\ and \dircrc\ have similar write performance when request size is small.
However, when request size increases, the overhead of CRC calculation dominates,
making \dircrc\ perform poorly.

We also evluated all the \dpm\ systems' write performance without read validation (\ie, treating \dpm\ as volatile memory).
We found each read validation to cost a constant of 1.5\mus\ overhead.

Figure~\ref{fig-dpm-read-latency} plots the average read latency.
Overall, \sepredo's performance is the best among \dpm\ systems and is only slightly worse than LITE.
However, when request size is small, \dircrc\ outperforms \sepredo\
because of \dircrc's read only requires one round trip under any circumstance.
However, like writes, the read performance of \dircrc\ dramastically drops as request size increases because of the CRC calculation overhead.
As expected, \dir\ and \ctrlredo\ perform worse than \sepredo\ because of their reads involve 3 RTTs and 2 RTTs.

\subsection{YCSB Results}

{
\begin{figure*}[th]
\begin{minipage}{1.3\columnwidth}
\begin{center}
\centerline{\includegraphics[width=1.0\columnwidth]{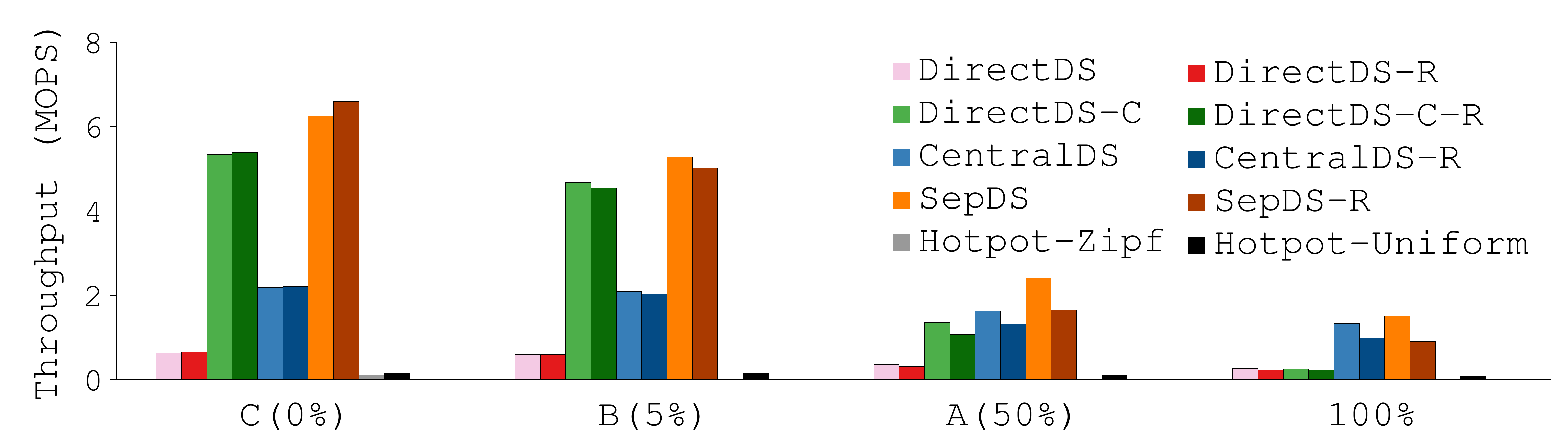}}
\vspace{-0.1in}
\mycaption{fig-dpm-replication}{\dpm{} Throughput and Replications}
{
Running YCSB on four \cn{}s and four \dpm{}s, with 1K request size and replication degree one and three. Hotpot is running without replication with four nodes and 8 threads per data node.
}
\end{center}
\end{minipage}
\begin{minipage}{0.1in}
\hspace{0.01in}
\end{minipage}
\begin{minipage}{0.67\columnwidth}
\begin{center}
\centerline{\includegraphics[width=0.9\columnwidth]{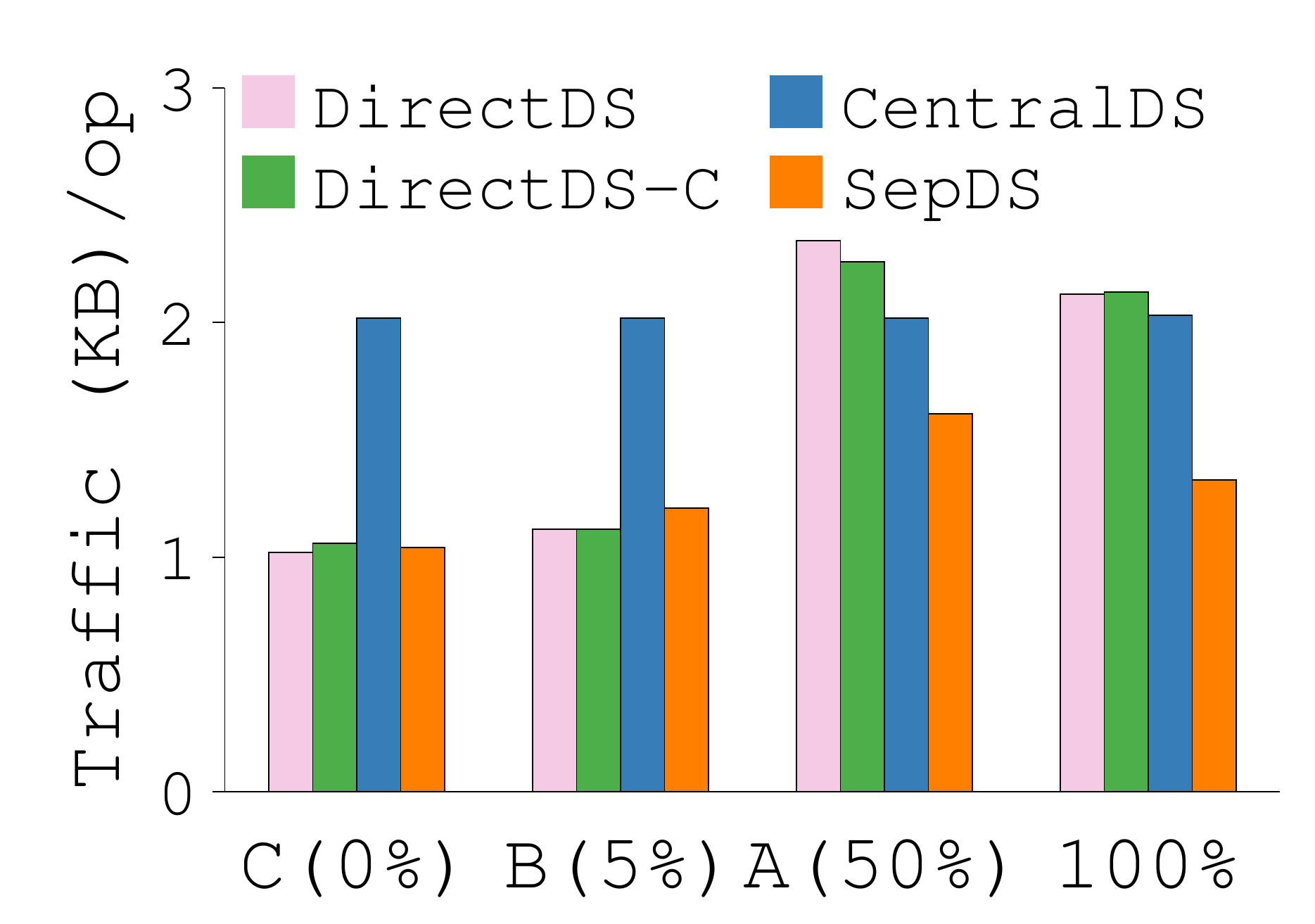}}
\vspace{-0.1in}
\mycaption{fig-dpm-network}{Network Traffic of \dpm\ stores}
{
Network traffic includes both control and data communication.
}
\end{center}
\end{minipage}
\end{figure*}
}

We now present our evaluation results using the YCSB benchmark~\cite{Cooper10-CloudCom,YCSB-C}.
We use a total of 100K key-value entries where the key size is 8 bytes and the value size is 1\KB.
The accesses to keys follow the Zipf distribution.
And we use four workloads with different read and write intensity:
read only (workload C), 
5\% write (workload B),
50\% write (workload A),
and write only.

\noindent{\textit{\uline{Basic performance.}}}
We first evaluate the performance of all our \dpm\ systems under 
our default setting: 4 \cn{}s and 4 \dpm{}s, each \cn\ runs 8 application threads.
Figure~\ref{fig-dpm-replication} shows the overall performance of \dpm\ data stores,
replicated \dpm\ data stores (with degree of replication 2), and Hotpot~\cite{Shan17-SOCC}.
The Hotpot runs use four servers, each running 8 application threads,
and we ran Hotpot with its MRSW (multiple reader single writer) consistency level without replication.
Hotpot serves as a comparison of the distributed \nvm\ model.

\sepredo\ performs the best among all systems regardless of read/write intensity,
even under high contention (with Zipf distribution to keys).
\dircrc\ performs well with workloads that are read intensive.
\dircrc's read performance is not affected by contention, since it does not need to perform any lock.
In contrast, \dirlock's read performance is the worst under contention because of it needs to lock a data entry for each read.
\ctrlredo's read performance is worse than \dircrc\ and \sepredo\
because \SY{each} read in \ctrlredo\ requires 2 RTTs and under contention the \coord\ becomes the bottleneck.

For write-intensive workloads, \ctrlredo\ and \sepredo\ perform better than the \dir\ systems.
This is because under high contention, the lock overhead of the \dir\ systems become high,
while \ctrlredo\ and \sepredo\ both avoid the lock contention.
\ctrlredo\ avoids it by using a local lock to protect metadata update (not the write itself)
and \sepredo\ uses the lock-free out-of-place chained data structure.

The overall performance of Hotpot is orders of magnitude worse than all \dpm\ data stores.
The reason is that each read and write in Hotpot involves a complex protocol that 
requires RPCs across multiple nodes.
Hotpot's performance is especially poor with writes, 
since the distributed \nvm\ consistency protocol involves frequent invalidation of cached copies,
especially under high write contention to the same data.
To confirm this, we also ran Hotpot with workloads with uniform distribution and found the results to be better, but still much worse than \dpm\ systems.

\noindent{\textit{\uline{Replication overhead.}}}
As expected, adding redundancy lowers the throughput of all data stores with write-heavy workloads.
Even though all systems issue the replication requests in parallel, 
they only use one thread to perform asynchronous RDMA read/write operations
and doing so still has an overhead.

\noindent{\textit{\uline{Network traffic.}}}
To further understand the cause of performance differences, 
we record the total network traffic during each run. 
Figure~\ref{fig-dpm-network} plots the average amount of traffic 
that each data store incurs to complete one operation under different workloads. 
\sepredo, \dirlock, and \dircrc\ send less traffic in read-heavy workloads
since these data stores access \dpm{}s directly.
In contrast, \ctrlredo\ incurs high traffic because data is sent once between the \cn\ and the \coord\ and once between the \coord\ and the \dpm.
As expected, \dir\ and \dircrc\ send more data for writes because of their writes involve 2 RTTs with data.


{
\begin{figure*}[th]
\begin{minipage}{0.97\columnwidth}
\begin{minipage}{0.48\columnwidth}
\begin{center}
\centerline{\subfloat[Workload B (5\%)]{\label{fig-dpm-mem-scalability-read}\includegraphics[width=1.0\columnwidth]{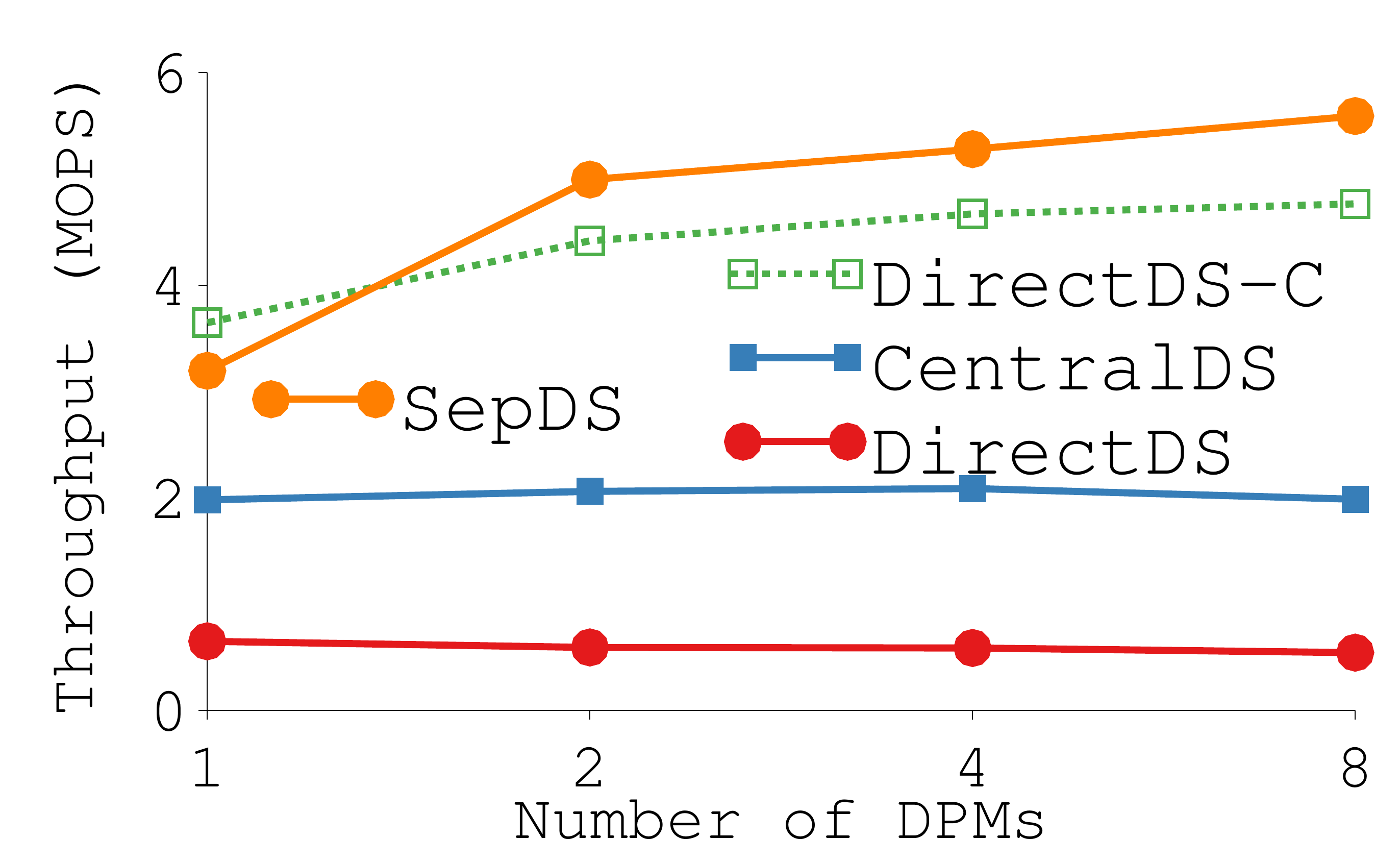}}}
\vspace{-0.1in}
\end{center}
\end{minipage}
\begin{minipage}{0.48\columnwidth}
\begin{center}
\centerline{\subfloat[Workload A (50\%)]{\label{fig-dpm-mem-scalability-write}\includegraphics[width=1.0\columnwidth]{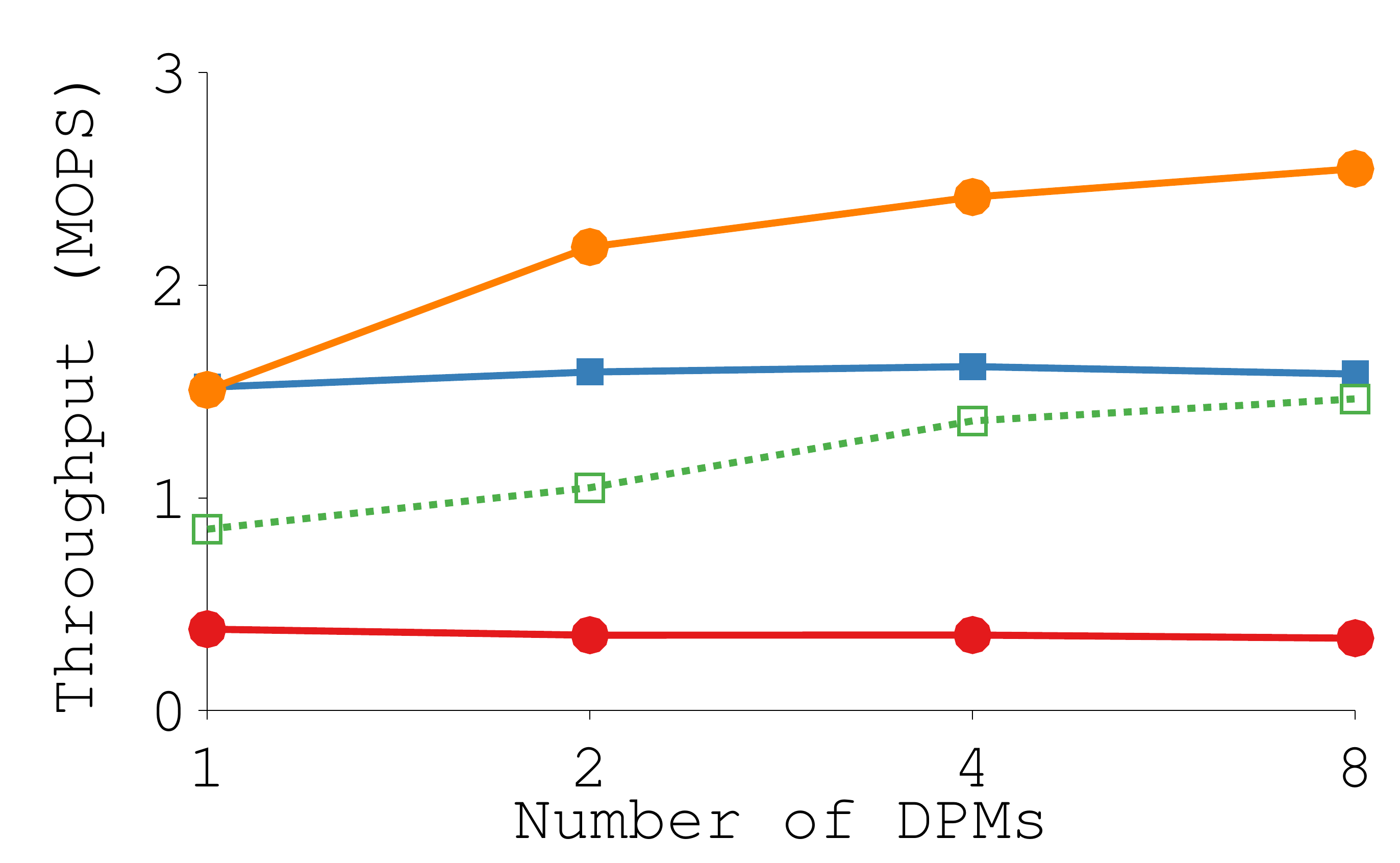}}}
\vspace{-0.1in}
\end{center}
\end{minipage}
\mycaption{fig-mem-scalability}{Scalability w.r.t. \dpm{}s}
{ 
Running 4 \cn{}s. Each \cn{} runs 8 threads.
}
\end{minipage}
\begin{minipage}{0.1in}
\hspace{0.01in}
\end{minipage}
\begin{minipage}{0.97\columnwidth}
\begin{minipage}{0.48\columnwidth}
\begin{center}
\centerline{\subfloat[Workload B (5\%)]{\label{fig-dpm-clt-scalability-read}\includegraphics[width=1.0\columnwidth]{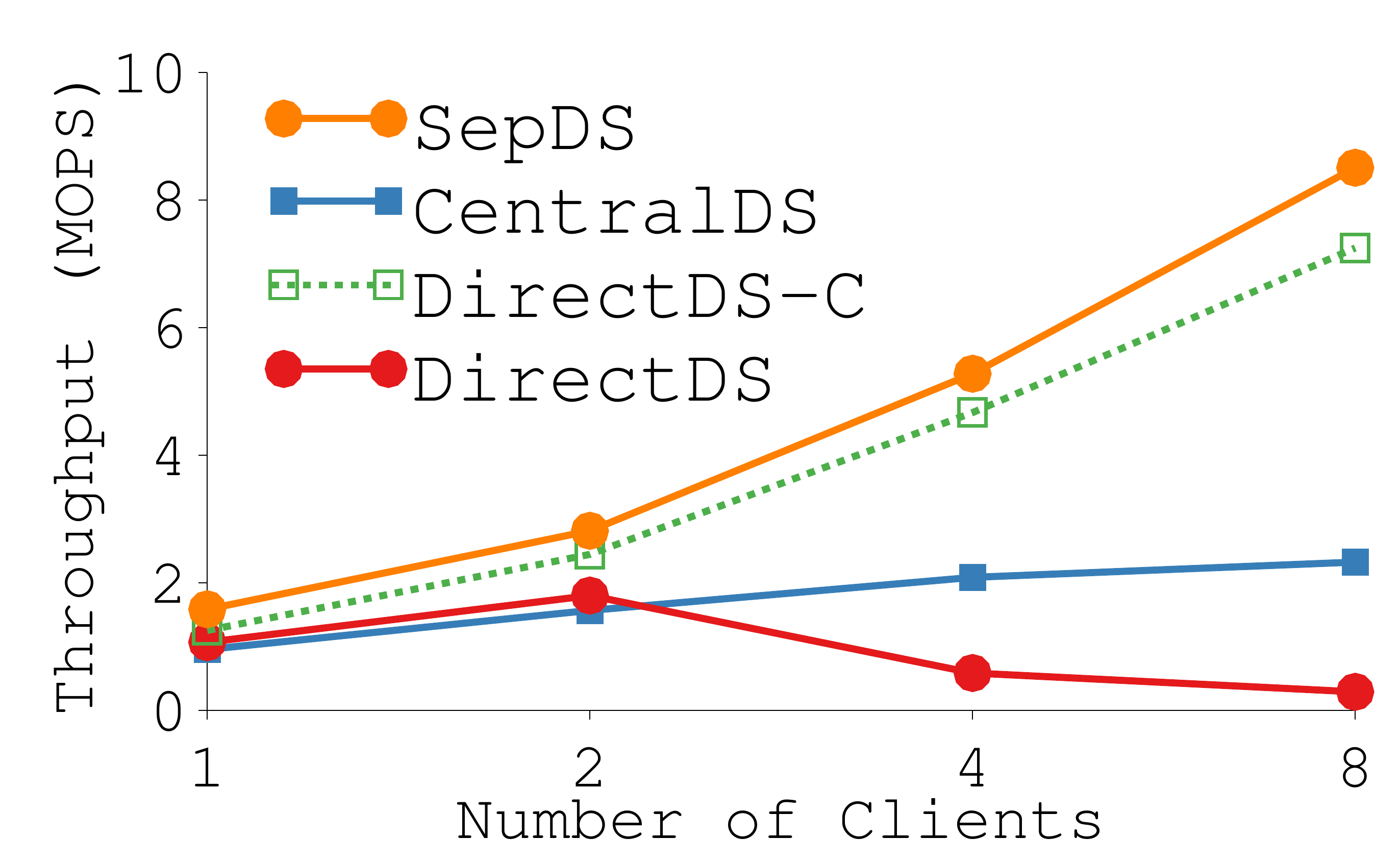}}}
\vspace{-0.1in}
\end{center}
\end{minipage}
\begin{minipage}{0.48\columnwidth}
\begin{center}
\centerline{\subfloat[Workload A (50\%)]{\label{fig-dpm-clt-scalability-write}\includegraphics[width=1.0\columnwidth]{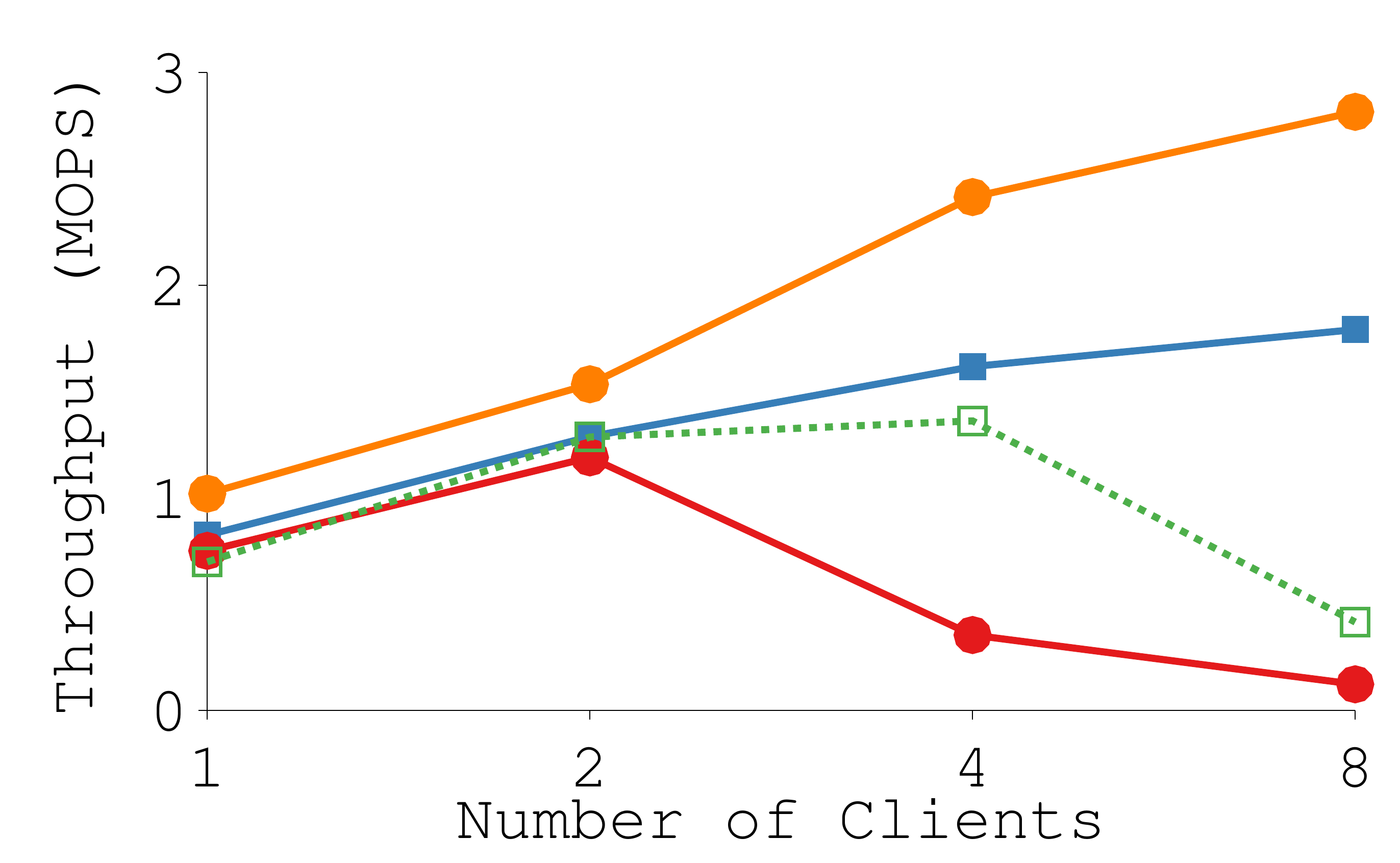}}}
\vspace{-0.1in}
\end{center}
\end{minipage}
\mycaption{fig-clt-scalability}{Scalability w.r.t. \cn{}s}
{ 
Running 4 \dpm{}s. Each \cn{}s runs 8 threads.
}
\end{minipage}
\end{figure*}
}
{
\begin{figure*}[th]
\begin{minipage}{0.66\columnwidth}
\begin{center}
\centerline{\includegraphics[width=1.0\columnwidth]{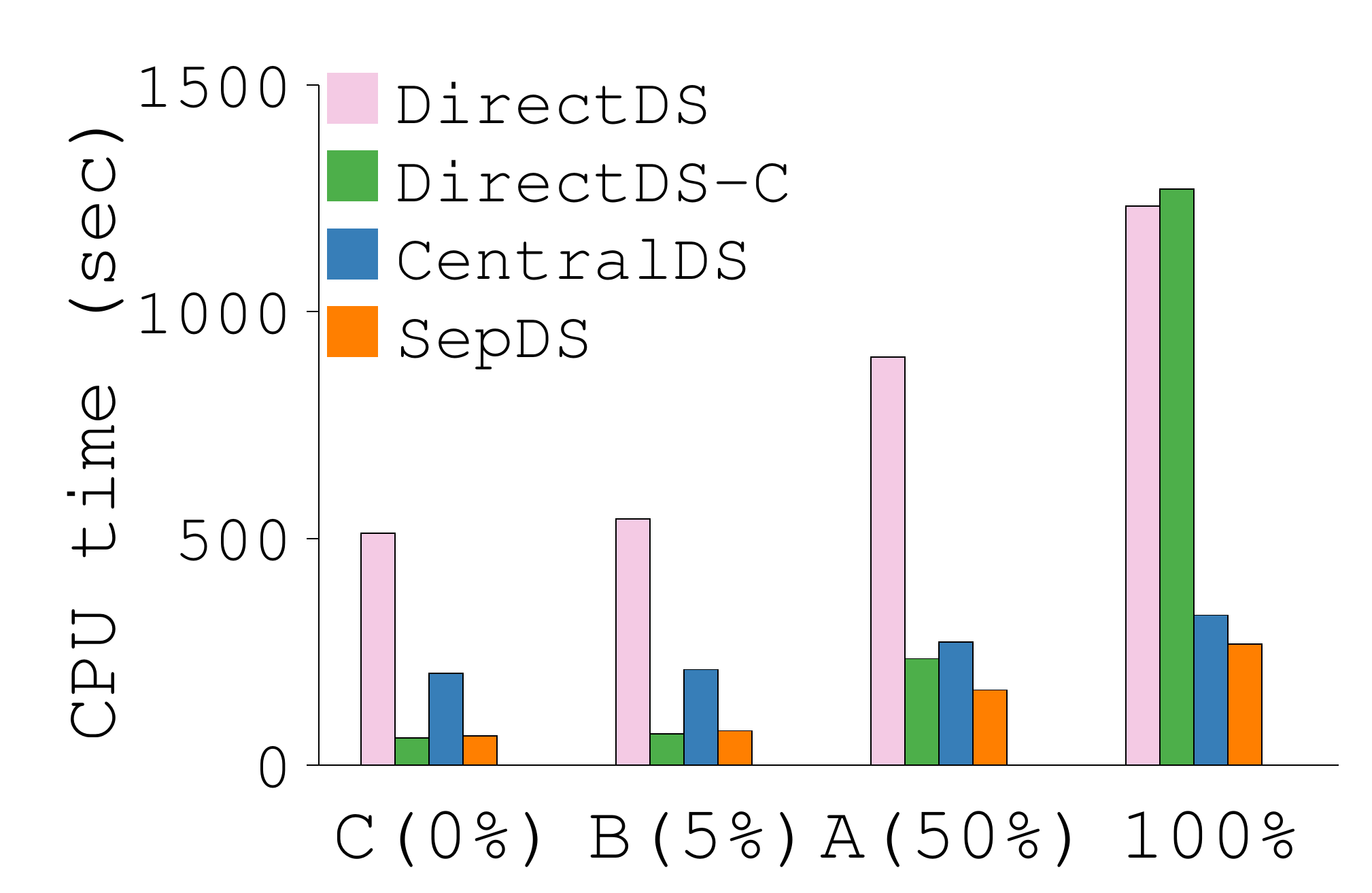}}
\vspace{-0.1in}
\mycaption{fig-dpm-CPU}{CPU Utilization}
{
CPU time to complete ten million requests. 
All tests run 4 \cn{}s each using 8 threads.
}
\end{center}
\end{minipage}
\begin{minipage}{0.1in}
\hspace{0.01in}
\end{minipage}
\begin{minipage}{0.42\columnwidth}
\begin{center}
\centerline{\includegraphics[width=1.0\columnwidth]{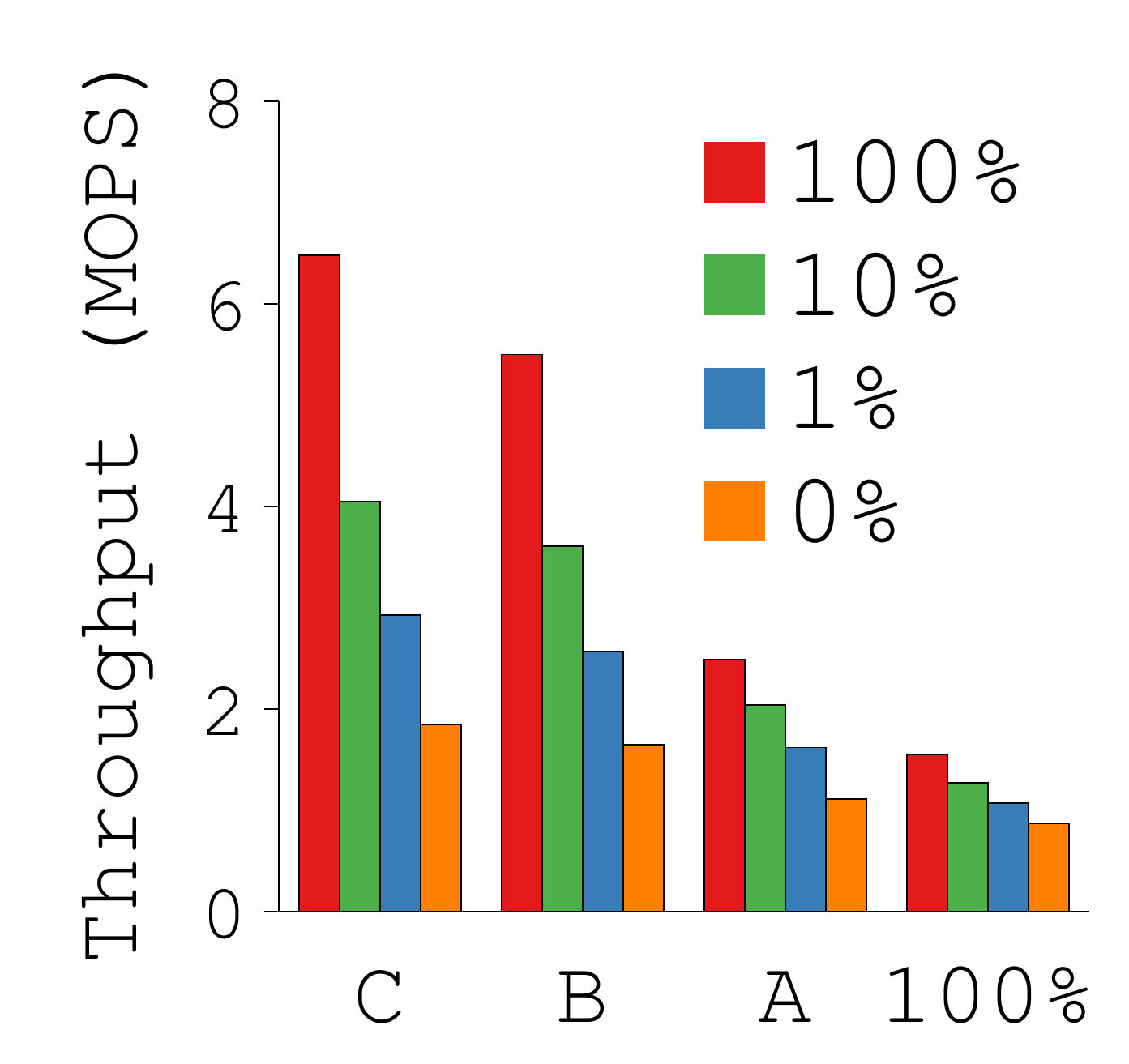}}
\vspace{-0.1in}
\mycaption{fig-dpm-cache}{Effect of Metadata Cache in \sepredo}
{
Each bar shows the percentage of total metadata each \cn{} can store.
}
\end{center}
\end{minipage}
\begin{minipage}{0.1in}
\hspace{0.01in}
\end{minipage}
\begin{minipage}{0.42\columnwidth}
\begin{center}
\centerline{\includegraphics[width=1.0\columnwidth]{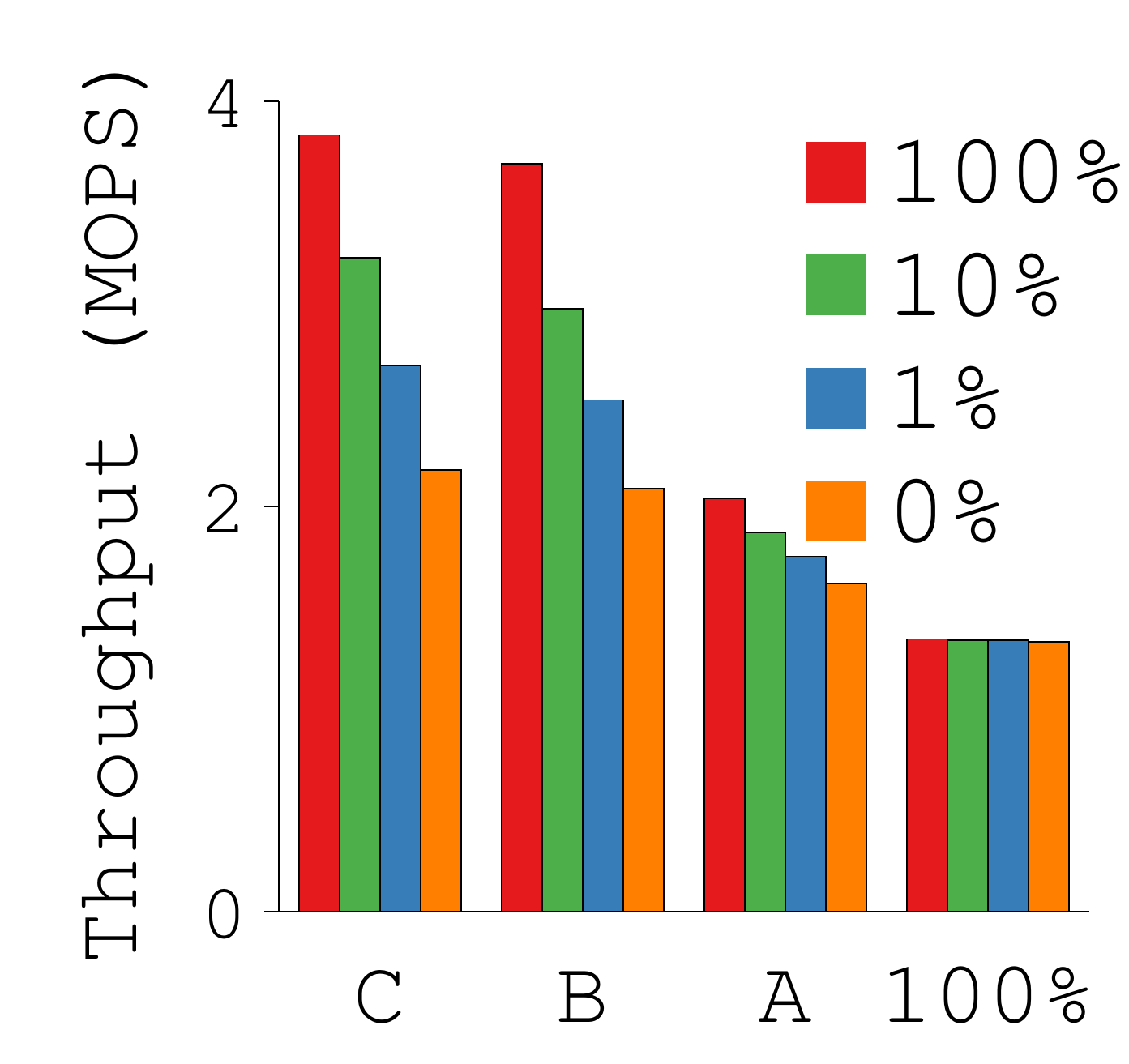}}
\vspace{-0.1in}
\mycaption{fig-dpm-central-cache}{Effect of Data Cache in \ctrlredo}
{
Each bar shows the percentage of total data the \coord\ can store.
}
\end{center}
\end{minipage}
\begin{minipage}{0.1in}
\hspace{0.01in}
\end{minipage}
\begin{minipage}{0.38\columnwidth}
\begin{center}
\centerline{\includegraphics[width=0.9\columnwidth]{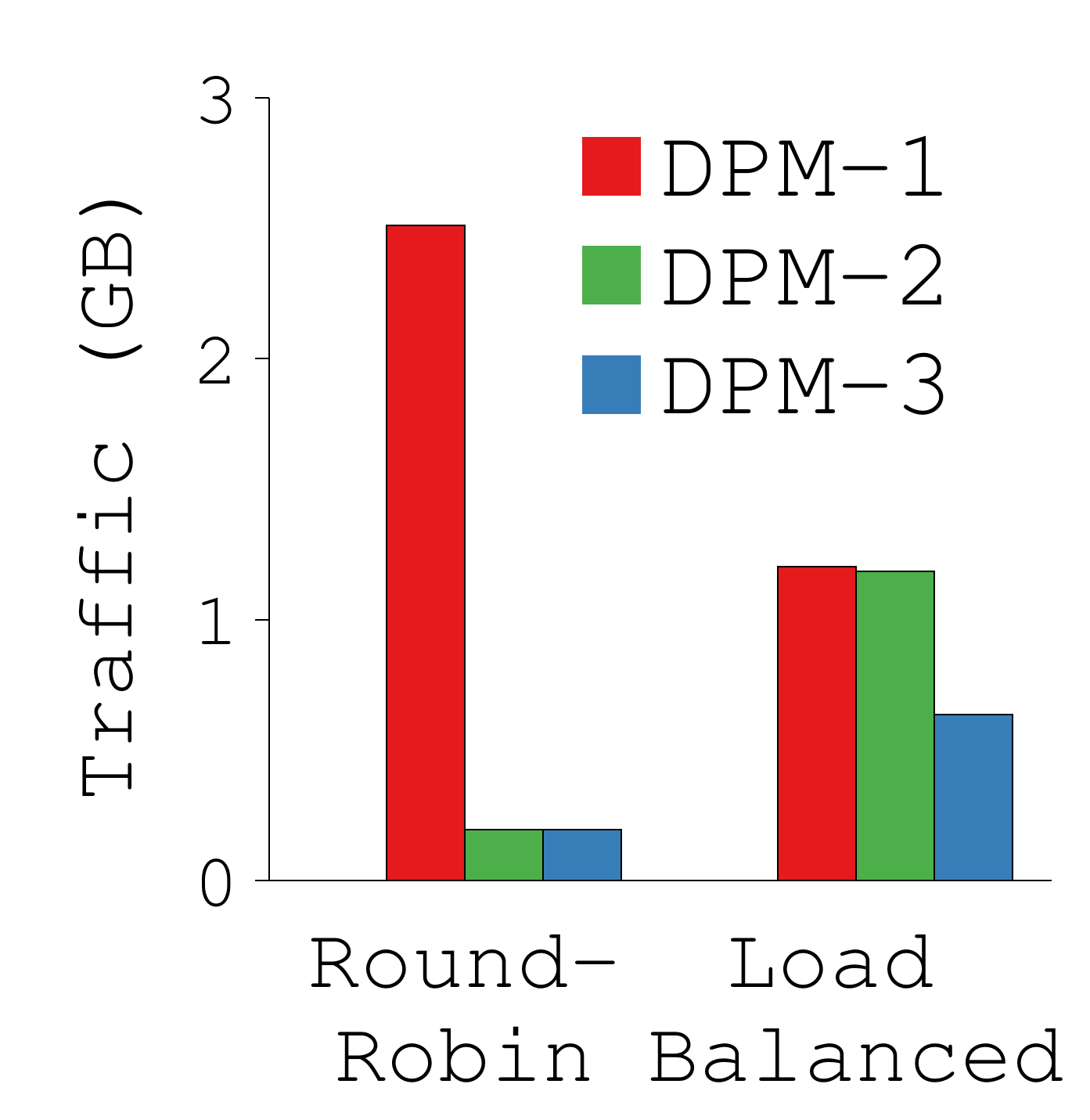}}
\vspace{-0.1in}
\mycaption{fig-dpm-load-balancing}{Load Balancing in \sepredo}
{
}
\end{center}
\end{minipage}
\vspace{-0.05in}
\end{figure*}
}

\noindent{\textit{\uline{Scalability.}}}
Next, we evaluate the scalability of different \dpm\ data stores with respect to the number of \cn{}s and the number of \dpm{}s.
Figure~\ref{fig-mem-scalability} shows the scalability of \dpm\ data stores w.r.t. the number of \dpm{}s.
Both \dircrc\ and \sepredo\ scale well with \dpm{}s because 
\dircrc\ and \sepredo\ both let \cn{}s access \dpm{}s directly,
improving the network bandwidth utilization to \dpm{}s.
\dirlock\ does not scale with \dpm{}s because of lock contention.
\ctrlredo\ does not scale well either, since its bottleneck is the network interface of a single coordinator.

Figure~\ref{fig-clt-scalability} shows the scalability of \dpm\ data stores 
when varying the number of \cn{}s.
\sepredo\ has the best scalability, since there is no single network bottleneck in \sepredo.
\ctrlredo's scalability is worse than \sepredo\
again because of the bottleneck of a single coordinator's network throughput.
\dirlock\ and \dircrc\ do not scale with \cn{}s.
As the number of \cn{}s increases, contention happens more frequently 
in \dirlock\ and \dircrc\ which reduce the overall throughput.

\noindent{\textit{\uline{CPU utilization.}}}
We evaluate the CPU utilization of different \dpm\ data store. 
Figure~\ref{fig-dpm-CPU} plots the total CPU time to complete ten million requests in different workloads.
For read-intensive workload, \dircrc\ and \sepredo\ use 
less CPU than other data stores because of one-sided primitives.
\dirlock\ suffers from lock contention which increases total CPU utilization.
For write-intensive workload, \sepredo\ uses less CPU time than 
other data stores mainly because \sepredo\ has higher 
throughput and separates data plane and control plane which reduces CPU usage.


\noindent{\textit{\uline{Metadata size.}}}
Different \dpm\ data stores cache different amounts of metadata in \cn{}s. 
\dirlock\ and \dircrc\ cache all keys and pointers to each entity for direct access to \dpm{}s.
\cn{}s in \ctrlredo\ only cache keys, and rely on coordinators to keep metadata.
\cn{}s in \dirlock\ and \dircrc\ keep the mapping from keys to \dpm{}s. 
Similarly, \sepredo\ caches a shortcut pointer for each entity to improve performance.
\sepredo\ further supports different sizes of metadata cache 

\noindent{\textit{\uline{Metadata caching effect.}}}
To evaluate the effect of different sizes of metadata cache at \cn{}s in \sepredo,
we ran the same YCSB workloads and configuration as 
Figure~\ref{fig-dpm-replication} and plot the results in Figure~\ref{fig-dpm-cache}.
Here, we use the FIFO eviction policy (we also tested LRU and found it to similar or worse than FIFO).
With smaller metadata cache, all workloads' performance drop because  
a \cn{} has to get the metadata from the \metaserver\ before accessing the data entry that does not have local metadata cache.
With no metadata cache (0\%), \cn{}s need to get metadata from the \metaserver\ before every request.
However, under Zipf distribution, with just 10\% metadata cache, \sepredo\ can already achieve satisfying performance.

\noindent{\textit{\uline{Data caching effect.}}}
We do not cache data at \cn{}s because doing so would require coherence traffic, resulting in performance that is similar to distributed \nvm.
However, it is possible to cache data at the \coord\ with the \scentral\ architecture, because that is the only copy and does not need any coherence traffic.
By caching hot data in a coordinator, the coordinator does not need 
to access \dpm{}s to get data for every read which can reduce 
network traffic and improve performance.
We built a FIFO data cache at the \coord\ for \ctrlredo\ to analyze the effect of data caching.
Figure~\ref{fig-dpm-central-cache} plots the throughput 
with different percentages of the data cache in a coordinator.
With bigger data cache, the performance increases.
However, the overall performance is still limited by network bandwidth.
Furthermore, we observe that data cache improves read traffic but not write traffic.
Overall, we found the effect of data caching to be small with \ctrlredo,
but demands large amount of \nvm\ space at the \coord.

\noindent{\textit{\uline{Load balancing.}}}
To evaluate the effect of \sepredo's load balancing mechanism,
we use a synthetic workload with three entities, $A$, $B$, and $C$.
We first create $A$ (without replication) and $B$ (with 2 replicas) 
and read these two entities heavily.
Then, we create $C$ (without replication) and keep updating $C$.
One \cn\ runs this synthetic workload on three \dpm{}s.
Figure~\ref{fig-dpm-load-balancing} shows the total traffic to 
the three \dpm{}s with and without load balancing.
With a naive allocation policy of round-robin across \dpm{}s, 
write traffic spreads among all \dpm{}s 
and read traffic only goes to the first \dpm.
With load balancing, \sepredo\ spreads read traffic across 
different replicas depending on the load of \dpm{}s.
At the same time, \metaserver\ allocates free entries 
for new writes from the least accessed \dpm.
As a result, the total loads across the three \dpm{}s are balanced.

\section{Related Work}
\label{sec:related}

Lim \etal~\cite{Lim09-disaggregate, Lim12-HPCA} first proposed the concept of disaggregating memory from processor.
Recent years have seen more industry and academic efforts in network support for disaggregated memory~\cite{OpenCAPI, Genz-citation, ccix-citation, Faraboschi-HotOS15, Novakovic14-ASPLOS} 
and software systems to manage remote memory~\cite{Dragojevic14-FaRM,Gu17-NSDI, Aguilera17-SoCC, Novakovic16-SoCC, Klimovic18-ATC}.
FaRM~\cite{Dragojevic14-FaRM,Dragojevic15-FaRM} is an RDMA-based distributed memory platform. 
FaRM use one-way communication for reads and perform both two-way and one-way communication for replicated writes (depending on whether it is to the primary copy).
Pilaf~\cite{Mitchell13-ATC} and HERD~\cite{Kalia14-RDMAKV, Kalia16-ATC} are two RDMA-based key-value store systems. 
These systems rely on two-way communication for writes and HERD and FaSST use two-way communication for reads too.

NAM-DB~\cite{Zamanian17-VLDB,Binnig16-VLDB} is a \rdma-based database system that uses one-sided communication for both read and write.
Infiniswap~\cite{Gu17-NSDI} is an RDMA-based remote memory paging system. 
Remote regions~\cite{Aguilera18-ATC} is a system that exposes 
remote memory as files that other host servers can access (through a file system interface). 
Although these three systems do not use two-way communication for data path, 
they both rely on processing power at remote nodes to run data management tasks.
\sepredo\ runs all management tasks (control path) at \metaserver, a separate node from remote memory.






Mojim~\cite{Zhang15-Mojim}, Hotpot~\cite{Shan17-SOCC}, and Octopus~\cite{Lu17-ATC} are three 
recent distributed \nvm\ systems. 
Mojim~\cite{Zhang15-Mojim} is the first system that targets 
using \nvm\ in distributed, datacenter environments. 
Mojim provides an efficient, RDMA-based, asynchronous replication 
mechanism for \nvm, to make it more reliable and available.
Hotpot~\cite{Shan17-SOCC} is the first distributed shared persistent memory system. 
It integrates the idea of distributed shared
memory and distributed storage systems to provide a 
globally coherent, crash-consistent, and reliable distributed \nvm\ system that 
applications can access with memory instructions.
Octopus~\cite{Lu17-ATC} is a distributed file system built on top of \nvm. 
None of these systems build on the \dpm\ model, which presents a whole new set of challenges.

ReFlex\cite{Klimovic17-ASPLOS} is a software-based system builds on IX~\cite{Belay14-OSDI} and exposes a logical block interface for users to access remote Flash with nearly identical performance as accessing local Flash. 
RAMCloud~\cite{RAMCLOUD} is a remote key-value storage system that stores a full copy of all data in DRAM and backups in disks or SSDs. 
Kamino-Tx~\cite{Memaripour17-EUROSYS} proposes a new mechanism to perform transactional 
updates on \nvm\ without any copying of data in the critical path.
These systems all rely on local computation power at remote memory/storage servers 
to perform various online and recovery management services which differs from \dpm\ model.
\section{Conclusion}
\label{sec:conclude}

This paper presents the disaggregated \nvm\ model, 
where \nvm\ is attached directly to the network without any local processors.
We proposed three \dpm\ architectures, built three 
atomic, crash-consistent, and reliable data stores on top of these architectures,
and performed extensive evaluation of these data stores.
Our findings will be able to guide future \dpm\ system builders.

\clearpage


\setstretch{0.8}
\titlespacing*{\section}{0em}{1ex}{1ex}
\begin{small}

\begin{spacing}{0.3}
\bibliographystyle{ACM-Reference-Format}
\bibliography{local}

\end{spacing}
\end{small}


\end{document}